\documentclass[aps,pre,amsmath,amsfonts,amssymb,superscriptaddress,nofootinbib]{revtex4}
\usepackage{graphicx}
\usepackage{psfrag}

\newcommand{\beq}{\begin{equation}} \newcommand{\eeq}{\end{equation}}

\def\stot{s_{\rm tot}}

\def\cG{{\cal G}}
\def\cA{{\cal A}}
\def\cV{{\cal V}}
\def\cE{{\cal E}}

\def\cH{{\cal H}}
\def\cHac{{\cal H}_{\rm ac}}
\def\cHsc{{\cal H}_{\rm sc}}
\def\cHpp{{\cal H}_{\rm pp}}

\def\W{\mathcal{W}}
\def\M{\mathcal{M}}
\def\tW{\widetilde{\mathcal{W}}}
\def\tM{\widetilde{\mathcal{M}}}

\def\lc{\lambda_{\rm c}}

\def\tF{\widetilde{F}}

\def\S{\mathcal{S}}
\def\hS{\widehat{\mathcal{S}}}

\def\hx{\widehat{x}}

\def\hsigma{\widehat{\sigma}}
\def\w{\omega}
\def\ow{\overline{\omega}}

\def\nM{\overline{M}}

\def\E{\mathbb{E}}
\def\I{\mathbb{I}}
\def\P{\mathbb{P}}

\def\T{\mathbb{T}}
\def\Z{\mathbb{Z}}
\def\tT{\widetilde{\mathbb{T}}}
\def\tu{\widetilde{u}}
\def\tg{\widetilde{g}}
\def\tc{\widetilde{c}}
\def\td{\widetilde{d}}
\def\dd{{\rm d}}
\def\di{\partial i}
\def\la{\langle}
\def\ra{\rangle}

\begin{document}

\title{Lifshitz tails on the Bethe lattice: a combinatorial approach}

\author{Victor Bapst}

\author{Guilhem Semerjian}
\affiliation{LPTENS, Unit\'e Mixte de Recherche (UMR 8549) du CNRS et
  de l'ENS, associ\'ee \`a l'UPMC Univ Paris 06, 24 Rue Lhomond, 75231
  Paris Cedex 05, France.}

\begin{abstract}

The density of states of disordered hopping models generically exhibits
an essential singularity around the edges of its support, known as a Lifshitz
tail. We study this phenomenon on the Bethe lattice, i.e. for the large-size
limit of random regular graphs, converging locally to the infinite regular 
tree, for both diagonal and off-diagonal disorder.
The exponential growth of the volume and surface of balls on these 
lattices is an obstacle for the techniques used to characterize the Lifshitz 
tails in the finite-dimensional case. We circumvent this difficulty by 
computing bounds on the moments of the density of states,
and by deriving their implications on the behavior of the integrated density
of states.

\end{abstract}

\maketitle

\section{Introduction}

The seminal work of Anderson~\cite{Anderson58} has given birth to a vast
body of literature on the properties of transport in random environments,
both in physics (see~\cite{Fifty} for a recent review) and in mathematics 
(monographs include for instance~\cite{carmona,PaFi,Stoll}). One central 
question in this domain is to determine whether a particle can diffuse
freely in the environment, according to the dimensionality of the model,
the intensity of the disorder, and the energy of the particle. This question
can be rephrased in terms of the spreading of eigenvectors of the Hamiltonian
corresponding to this energy, or more mathematically in terms of the nature
(i.e. absolutely continuous versus pure-point) of its spectrum.
Another direction of investigation of these models concerns the density of
states~\cite{IDS} of such random Hamiltonians, roughly speaking the 
distribution of eigenvalues, irrespectively of the nature (localized or 
extended) of the corresponding eigenvectors. Even though this quantity does
not reflect directly the localization properties of the 
Hamiltonian~\cite{Wegner,Simon83}, its study by Lifshitz~\cite{Li64} revealed 
very early an interesting behavior around the edge of the spectrum, namely
a very fast vanishing following an essential singularity known as a Lifshitz 
tail. This behavior can be explained intuitively as follows. In dimension $d$,
an eigenvector of the non-disordered Hamiltonian with an eigenvalue close
to the band edge, say at a very small distance $\delta$ from it, is supported
on a volume of order $\delta^{-d/2}$. In presence of disorder of bounded 
amplitude, this vector
can give rise to an eigenvalue at an energy $\delta$ lower than the (displaced)
band edge $E_{\rm max}$ only if the random potentials on all sites in this 
volume are close to their maximal value. The probability of this event is 
exponentially small in the number of involved sites, hence the form of the 
Lifhsitz tail in  dimension $d$, 
$\rho(E_{\rm max}-\delta) \approx \exp[- \delta^{-d/2}]$. This heuristic 
reasoning can be turned into a rigorous derivation, see for instance
chapter VI.2 in~\cite{carmona} and~\cite{Simon85} for an illuminating 
exposition. Note also that even if the density of states is not directly
related to the localization aspects of the problem, proofs of localization
in finite dimension based on the multi-scale analysis~\cite{FS83} rely 
crucially on its estimates.

It has been realized early on~\cite{ACTA73,ACT74} that the Anderson model
could also be studied on the Bethe lattice, thus enabling a mean-field analysis
of the localization transition (that does not exist on the fully-connected, 
complete graph). 
Since then this version of the model has been the subject of several
works both in physics~\cite{KiHa85,MiFy91,MiDe94,BAF04,MoGa09,BST10} 
and in mathematics~\cite{KuSo83,AcKl92,AM93,Kl98,ASW06,FHS07,AW10,AW11}. 
Among other results these papers contain numerical procedures
to compute the density of states and the location of the
localization transition~\cite{ACTA73,ACT74,MiDe94,BAF04,MoGa09,BST10}, 
as well as proofs of the existence of an absolutely continuous part of the 
spectrum at low disorder~\cite{Kl98,ASW06,FHS07},
and of localization at large disorder or on the border of the 
spectrum~\cite{AM93}. The study of sparse random 
matrices~\cite{RoBr88,BiMo99,BaGo01,SeCu02,KhShVe04,RoCaKuTa08,MeNeBo10} is
actually closely connected to the Anderson problem on the Bethe lattice,
even though the perspective taken is slightly different; in the latter case
the disorder appears through the connectivity properties of the underlying 
random graph with fluctuating degrees.

\begin{figure}[h]
\center
\includegraphics[height=6.5cm]{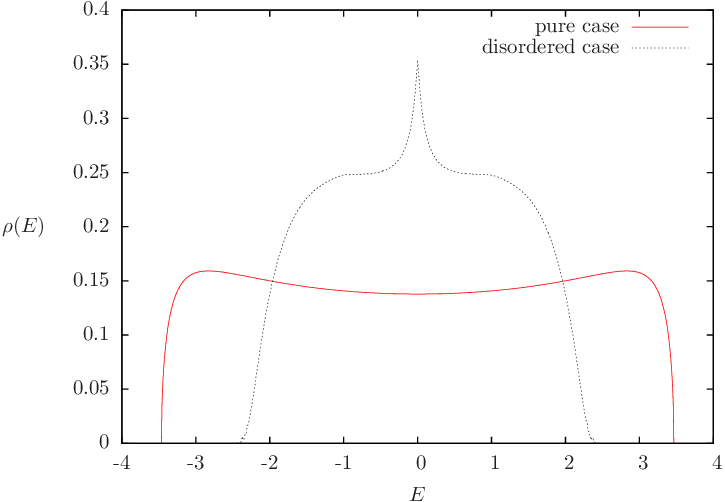} \hspace{1 cm}
\caption{Density of states on the Bethe lattice of degree $k+1=4$, 
in the pure ($J_{ij}=1$) and disordered case (with off-diagonal disorder 
$J_{ij}$ uniformly distributed on $[-1,1]$, see below for a precise 
definition). The pure case result
is the Kesten-McKay~\cite{kesten,mckay} density given in 
Eq.~(\ref{eq_density_pure}). The curve of the disordered case 
was obtained following the numerical procedure recalled in 
Appendix~\ref{appendix_middle}. The support of these two distributions
is the same, despite the apparent vanishing of the disordered one due
to the strong Lifshitz tail effect.}
\label{fig_numerical_simulation}
\end{figure}

We shall focus in this paper on the Lifshitz tail phenomenon for the Bethe
lattice geometry. Heuristic arguments put forward for instance 
in~\cite{KiHa85,BST10} and a rigorous analysis for a particular
type of disorder~\cite{muller,reinhold} suggest an even more violent behavior 
of the density of states in this regime, namely a vanishing of the form 
$\rho(E_{\rm max}-\delta) \approx \exp[-\exp[\delta^{-1/2}]]$, that makes this 
regime very hard to study numerically. As a first example we show in
Fig.~\ref{fig_numerical_simulation} a plot of the density of states obtained
by a standard numerical procedure recalled in 
Appendix~\ref{appendix_middle}, which displays an apparent band edge
far from its exactly known value. This different form (doubly-exponential) 
of the Lifshitz tail
with respect to the finite-dimensional case can be associated to
the exponential (instead of polynomial) growth of the volume of a ball as a
function of its radius on the Bethe lattice (formally corresponding to
$d\to \infty$). In addition the surface of a ball is asymptotically equivalent
to its volume on such a non-amenable graph. This strongly complicates the
transposition of the scheme of proofs of Lifshitz tails from the finite-$d$ 
case to the Bethe lattice one, and indeed rigorous results on the Lifshitz 
tail behavior on Bethe lattices or sparse random graphs are restricted to 
sub-critical percolation models \cite{KhKiMu06,muller,reinhold}.
The alternative method developed in this paper to circumvent this difficulty 
consists in studying the asymptotic behavior of the moments of the density of 
states. The bounds on the moments that we obtain, and their consequences
for the integrated density of states, are in agreement with the
doubly-exponential form of the Lifshitz tail on the Bethe 
lattice~\cite{KiHa85,BST10,muller,reinhold}.

The rest of the paper is organized as follows. 
In Sec.~\ref{sec_def_and_results} we define more precisely the models under 
study and we state our main results. In Sec.~\ref{sec_expressions} we collect
several expressions of the moments of the density of states. The following two
sections (\ref{sec_bounds_offd} and \ref{sec_bounds_d}) present the proofs
of our results for the off-diagonal and diagonal disorder case, respectively.
Each of these two sections is divided into two subsections focusing on lower
and upper bounds on the moments of the density of states. Finally we draw our 
conclusions and propose perspectives for future work 
in Sec.~\ref{sec_conclusions}. 
More technical aspects of our work are collected in a series of 
Appendices.

\section{Definitions and results}
\label{sec_def_and_results}

In this section we shall define the density of states of Anderson models
on the Bethe lattice, and state our main results on the asymptotic growth
of its moments and their counterparts on the decay of the integrated density of states near the edge of the spectrum. The (integrated) density of states is usually defined, for
finite-dimensional models considered on the $\Z^d$ lattice, via a limiting
procedure from finite boxes $[-L,L]^d$ with $L \to \infty$. The corresponding
construction on an infinite regular tree, namely cutting a depth $L$ 
neighborhood around an arbitrarily chosen vertex, is not satisfactory
because the number of vertices on the surface of a ball of radius $L$ is of
the same order as the one of its interior. Finite-size regularizations of
the Bethe lattice are instead provided by random regular graphs (see for 
instance~\cite{MePa_Bethe} for such a discussion in the context of 
spin-glasses). We shall hence begin this section by defining both
finite (in Sec.~\ref{sec_def}) and infinite 
(in Sec.~\ref{sec_def_model_infinite}) versions of the Bethe lattice models, 
and highlight the connection between the two. 
We will in particular define the density
of states and its moments, and discuss its regularity properties in 
Sec.\ref{subsection_regularity}. Our main results on the asymptotic growth of
the moments of the density of states are given in~\ref{sec_res_bounds};
to make them more intuitively understandable we discuss
just before, in Sec.~\ref{sec_density_to_moments}, how the behavior of a
probability density close to its edge is reflected in the type of growth
of its moments.

\subsection{Anderson models on random regular graphs} 
\label{sec_def} 

Let us consider a finite graph $\cG=(\cV,\cE)$ on $N$ vertices, and a 
$N\times N$ symmetric matrix $H$ defined by its matrix elements
\begin{equation}
H_{ij} = \begin{cases}
V_i & \text{if} \ i=j \\
J_{ij} & \text{if} \ (i,j) \in \cE \\
0 & \text{otherwise}
\end{cases} \ .
\end{equation}
The real numbers $V_i$ (resp. $J_{ij}=J_{ji}$) represent the influence of the
disorder on the vertex $i$ (resp. on the edge between $i$ and $j$). 
In the pure case (without disorder), i.e. when $V_i=0$ and $J_{ij}=1$ for
all vertices and edges, $H$ is nothing but the adjacency matrix of the graph.
This well-known case will be considered for comparison purposes in the 
following.

For a given realization of the Hamiltonian matrix $H$ its empirical spectral 
measure is defined as a sum of Dirac atoms on its eigenvalues 
$\lambda_1,\dots,\lambda_N$:
\begin{equation} 
\widehat{\rho}_N = \frac{1}{N}\sum_{i=1}^N \delta_{\lambda_i} \ .
\end{equation}

We turn this measure into a random object by taking for $\cG$ a random 
$k+1$-regular graph (i.e. choosing it uniformly at random among all graphs
on $N$ vertices, where all vertices have degree $k+1$), for the diagonal
elements $V_i$ a set of independent identically distributed (i.i.d.) random
variables, and another set of i.i.d. random variables for the $J_{ij}$'s with
$i<j$. The probability measure associated to this construction will be
denoted $\P[\cdot]$, and corresponding averages as $\E[\cdot]$.
In the following we will concentrate mostly on two particular cases: 
\begin{itemize}
\item the off-diagonal (or bond) disorder case, in which all the $V_i$ 
vanishes.
\item the diagonal (or site) disorder case, in which all the $J_{ij}$ are 
equal to $1$.
\end{itemize}
In the former case we shall make the following assumptions on the
random edge couplings:
\begin{itemize} 
\item (OD1) $J$ has a support included in $[-1,1]$, with $|J|$ not always
equal to $1$, but which takes values arbitrarily close to $1$ with
positive probability.
In technical terms, we assume the existence of 
$J_{\rm min} > 0$ such that $0<\P[|J|\ge J_0] <1$ for all
$J_0$ with $J_{\rm min} \le J_0 <1$.
\item (OD2) $\forall \alpha>0 \ , \; a \in (0,1/2) \ , \;
|\ln  \P(J\ge 1-\epsilon)|e^{-\alpha/\epsilon^a} = 
O(\epsilon)$ when $\epsilon \to 0$. 
\end{itemize}
The corresponding assumptions for the random energies $V_i$ in the latter case
will be:
\begin{itemize} 
\item (D1) $V$ has a support included in $[0,W]$, distinct from $\{W\}$,
giving positive weight to a neighborhood of $W$.
\item (D2) $\forall \alpha>0 \ , \; a \in (0,1/2) \ , \;
|\ln  \P(V\ge W-\epsilon)|e^{-\alpha/\epsilon^a} = 
O(\epsilon)$ when $\epsilon \to 0$.
\end{itemize}
The simplest examples of random variables satisfying
these assumptions are the uniformly random ones (on $[-1,1]$ for $J$ and
on $[0,W]$ for $V$), or the Bernoulli random variable which corresponds,
in the off-diagonal case, to the bond percolation model on the Bethe lattice.
The assumptions (OD2) and (D2) imply that the occurrence
of a $J_{ij}$ (or $V_i$) close to its maximal value is not by itself a very
rare event, otherwise the Lifshitz tail phenomenon would be controlled by 
single site large deviations and would not be a collective effect anymore,
as will become clear in Sec.~\ref{sec_bounds_offd_lower} and 
Sec.~\ref{sec_bounds_d_lower}.
Note that the assumption (D1) does not hold for the Cauchy and Gaussian 
distributions of diagonal disorder studied for instance 
in~\cite{AcKl92,MiDe94,AW10,AW11}. The choice of a positive support for the diagonal disorder will greatly simplify the proofs, without loss of generality: adding a 
constant value to the $V_i$'s only shifts the support of $\widehat{\rho}_N$.

The local convergence of random regular graphs to infinite trees 
ensures~\cite{lelarge} that $\widehat{\rho}_N$ 
converges (weakly, in distribution) in the $N\to \infty$ limit to a 
probability measure $\rho$, that we shall call, following the physics
convention with a slightly abusive terminology discussed in the next 
subsections, the density of states.
For instance in the pure case where $H$ is the adjacency matrix of a
random $(k+1)$-regular graph, $\rho$ is the Kesten-McKay 
measure~\cite{kesten,mckay}, supported on $[-2\sqrt{k},2\sqrt{k}]$ with the 
density
\begin{equation}
\rho_{\rm KM}(E) = \frac{k+1}{2\pi} \frac{\sqrt{4 k - E^2}}{(k+1)^2-E^2} \ .
\label{eq_density_pure}
\end{equation}
In the next subsection we shall see how to define the measure $\rho$ 
directly on an infinite object without this $N\to\infty$ limit.
Before that we define the average 
moments of the empirical spectral measure,
\begin{equation}
u_{n,N} \equiv \E \left[ \int \lambda^n \dd\widehat{\rho}_N(\lambda)\right]
= \frac{1}{N} \E \left[ \sum_{i=1}^N \lambda_i^n \right] 
= \frac{1}{N} \E \left[ \textrm{tr} H^n \right] 
= \E \left[ (H^n)_{00} \right] \ ,
\label{eq_def_unN}
\end{equation}
where in the last step we have used the invariance of the random graph 
ensemble with respect to the permutations of the vertices and denoted $0$ the
index of an arbitrary vertex in $V$. We shall use the symbol $u_{n,N}$ for
a generic disorder distribution, and denote instead these average moments 
$c_{n,N}$ (resp. $d_{n,N}$) for the off-diagonal (resp. diagonal) disorder
case. We define $u_n$ (resp. $c_n$, $d_n$) as the $N\to \infty$
limit of $u_{n,N}$ (resp. $c_{n,N}$, $d_{n,N}$). The above stated convergence
of $\widehat{\rho}_N$ imply that this limit exists and that the $u_n$ are
the moments of the limiting measure, the density of states $\rho$ 
(a ``self-averaging'' property). The existence of the limit for the average
moments can in fact be obtained in a more direct way: expanding $(H^n)_{00}$
in the last expression of Eq.~(\ref{eq_def_unN}) shows that $u_{n,N}$ depends
only on the disorder in the subgraph neighboring the reference vertex $0$
within a distance at most $n/2$, which converges in the $N\to \infty$ limit
to a regular tree.

\subsection{Anderson models on infinite trees}
\label{sec_def_model_infinite}

It is also possible to define Anderson models directly on infinite graphs, 
that we keep denoting $\cG=(\cV,\cE)$. 
The infinite counterpart of the matrix $H$
becomes an operator acting on the elements $\varphi$ of the Hilbert space
$\cH = \ell^2(\cV)$ as
\begin{equation}
(H\varphi)_i = \sum_{j \in \di} J_{ij} \, \varphi_j + V_i \, \varphi_i \ ,
\label{eq_def_H_infinite}
\end{equation}
where $\di$ stands for the set of neighbors of $i$ in the graph $\cG$.
When the variables $J_{ij}=J_{ji}$ on the edges
of $\cE$, the $V_i$'s on the vertices, and the degrees $|\di|$ of the vertices, 
are all uniformly bounded, then $H$ is a self-adjoint operator.
For the convenience of the physicist reader who wants to immerse him/herself 
in the 
mathematical literature we shall sketch in an informal way some of the 
concepts used in the mathematical literature on localization, even though 
this is not the main subject of the paper. Exhaustive reviews of this subject
can be found in the monographs~\cite{carmona,PaFi,Stoll} and in the lecture 
notes~\cite{invitation}.

A major difficulty in dealing with infinite dimensional Hilbert space is
that some of the solutions of the eigenvector equation $H \varphi = E \varphi$
are not normalizable, i.e. not in $\ell^2(\cV)$. The spectrum $\sigma(H)$
has thus to be defined as the complement of the resolvent set. The latter
is the set of complex numbers $z$ such that the resolvent operator (or Green
function) $(H-z\I)^{-1}$ exists and is bounded ($\I$ denotes the identity
operator). When $H$ is self-adjoint
$\sigma(H) \subset \mathbb{R}$. The spectral theorem~\cite{reedsimon} is the 
extension to infinite dimensional self-adjoint operators of the 
diagonalization of Hermitian matrices. It asserts that $H$ can be written
as
\begin{equation}
H = \int \lambda \; \dd \mu(\lambda) \ ,
\label{eq_def_spectral}
\end{equation}
where $\mu$ is a projection valued measure providing a resolution of
identity: for all (Borel) subsets $I$ of
$\mathbb{R}$, $\mu(I)$ is an orthogonal projection. In the finite 
dimensional case $\mu(I)$
would project on the subspace spanned by eigenvectors associated to
eigenvalues $E \in I$. From this projection valued measure one can define
the spectral measures, which are real measures associated to elements 
$\varphi \in \cH $ according to 
$\mu_\varphi(I) = \la \varphi|\mu(I)|\varphi \ra$. Any real measure $\eta$
can be decomposed~\cite{feller2} in three contributions of different types, 
$\eta = \eta_{\rm pp} + \eta_{\rm ac} + \eta_{\rm sc}$, where $\eta_{\rm pp}$
is the pure point part of $\eta$ (its set of Dirac peaks), $\eta_{\rm ac}$
its absolutely continuous part (which gives no weight to sets of zero
Lebesgue measure, and has a density thanks to the Radon-Nikodym theorem), and
the remainder $\eta_{\rm sc}$ is the singular continuous contribution. 
Combining this decomposition of real measures with the definition of the
spectral measures $\mu_\varphi$ leads to a partition of the Hilbert space
as $\cH = \cHac \oplus \cHpp \oplus \cHsc$, where $\cH_i$ is defined,
for $i=$ pp, ac, sc, as 
$\cH_i =\{\varphi \in \cH \, |\, \mu_\varphi \, \text{is purely of type}\ i \}$.
Each of these subspaces is stable under the action of $H$, one can thus
define the three spectra $\sigma_i(H)$ as 
$\sigma(H \upharpoonright \cH_i)$, where 
$H \upharpoonright \cH_i$ denotes the restriction of $H$ to $\cH_i$.
In general $\sigma_{\rm pp}(H)$, $\sigma_{\rm ac}(H)$ and $\sigma_{\rm sc}(H)$
are not disjoint. The pure-point spectrum contains the ``true'' eigenvalues,
corresponding to localized (normalizable) states, while the absolutely 
continuous spectrum is associated to generalized eigenvalues and extended 
states. This classification is reflected in the dynamical evolution of the
system: a wavepacket constructed from states in the absolutely continuous
spectrum spread out at infinity, while pure-point ones remain localized in
a finite volume.

This discussion concerned a given operator $H$; when the coupling
constants $J_{ij}$ and on-site energies $V_i$ are turned in random variables,
all the quantities defined above become themselves random. However when the
$J_{ij}$ and $V_i$ are independent (or more generally form an ergodic process),
it follows from ergodicity that the spectrum $\sigma(H)$ is equal,  with 
probability 1, to a deterministic set independent of the random variables 
$J_{ij}$ and $V_i$. Moreover this set is equal to the spectrum of a (pure) model
where all random variables are fixed deterministically to their extremal
values. A stronger ergodicity statement is actually true:
the spectra $\sigma_{\rm pp}(H)$, $\sigma_{\rm ac}(H)$ and $\sigma_{\rm sc}(H)$
are also independent, with probability 1, of the actual realization of the
disorder. One of the main goal of the localization
studies is the characterization of these sets, depending on the underlying
graph and the distribution of the disorder. In unidimensional models the
spectrum is completely pure-point for infinitesimal amounts of diagonal 
disorder~\cite{KuSo80}, while in large dimensions there exists a mobility edge 
separating the extended and localized~\cite{FS83,AM93} parts of the spectrum.

The ``density of states'' $\rho$ is defined as the average of the spectral 
measure associated to an element $\varphi \in \cH$ supported on a single 
arbitrary site $0 \in \cV$, $\varphi_i=\delta_{i,0}$, that we shall denote
\begin{equation}
\rho = \E [\la 0 | \mu | 0 \ra ] \ .
\label{eq_rho_def_infinite}
\end{equation}
Its denomination is a bit misleading from a mathematical point of view:
$\rho$ is a probability measure, a priori not absolutely continuous and
in consequence not associated to a density (we shall come back to this point
shortly afterwards). The mathematical literature calls instead integrated
density of states (IDS) the cumulative distribution function 
$N(E) = \rho((-\infty,E])$ which always exists.
In finite dimensions (i.e. for $\cV=\mathbb{Z}^d$) the IDS can be constructed
as follows: call $H^L$ the restriction of the operator $H$ to a box of
volume $L^d$ with some boundary conditions. $H^L$ acts on a finite-dimensional
space and can thus be diagonalized. Let $N^L(E)$ be the fraction of its
eigenvalues smaller than or equal to $E$; it can then be proven that $N^L$ 
converges as $L$ is sent to infinity to the (deterministic) distribution 
function $N$ defined above directly from the infinite-size
operator. Another general result on the IDS is that its support coincides
with the (almost sure) spectrum $\sigma(H)$.

\begin{figure}[h]
\center
\includegraphics{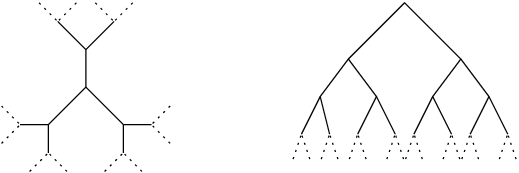}
\caption{Left: The regular rooted $3$-tree $\mathbb{T}_2$ - 
Right: The rooted $2$-ary tree $\widetilde{\mathbb{T}}_2$.}
\label{fig_trees}
\end{figure}

Let us now come back to the main subject of the paper and specialize some of 
the definitions above to the (infinite) Bethe lattice. In that case 
$\cG$ is the regular tree in which every vertex has $k+1$ neighbors, 
sketched in the left panel of Fig.~\ref{fig_trees} and denoted $\T_k$ in the
following. We shall also use the rooted
$k$-ary tree $\tT_k$, in which there is a distinguished vertex (the root) that 
has no parent node, and every vertex has exactly $k$ children, see right panel
of Fig.~\ref{fig_trees}. We will sometimes also consider $\T_k$ as a rooted
tree, choosing an arbitrary vertex $0$ as the root, and calling children of
a vertex $v \neq 0$ its $k$ neighbors away from the root.
We take the $J_{ij}=J_{ji}$ as a collection of i.i.d. random variables on the 
edges, and similarly with the $V_i$'s another set of i.i.d. random variables.
In the pure case ($J_{ij}=1$, $V_i=0$) the spectrum of $H$ is the interval
$[-2\sqrt{k},2\sqrt{k}]$. In the diagonal disorder case ($J_{ij}=1$) with the $V_i$ 
i.i.d. random variables supported on $[0,W]$ the ergodicity result mentioned
above (see~\cite{AcKl92} for a discussion of ergodicity on the Bethe lattice)
implies that the (almost sure) spectrum of $H$ and the support of $\rho$ is
$[-2\sqrt{k},2\sqrt{k}+W]$. Similarly in presence of off-diagonal disorder
($J_{ij}$ satisfying (OD1), $V_i=0$) the support of $\rho$ is the same as
in the pure case, $[-2\sqrt{k},2\sqrt{k}]$. Trees are in some respects much
simpler than finite-dimensional lattices, as they admit a natural recursive 
decomposition. One can for instance picture $\T_k$ as made of its root,
$k+1$ edges from the root to its neighbors, and $k+1$ copies of the rooted
$k$-ary tree $\tT_k$. This decomposition allows to write explicit recursion
relations on the diagonal elements of the resolvent (Green function). Let us
call $G(z)= \la 0 | (H-z\I)^{-1}|0 \ra$ the Green function at the root of
$\T_k$, where $z$ has a positive imaginary part. This is a random variable 
because of the disorder in the $J_{ij}$ and $V_i$. Using the recursive
nature of $\T_k$ and resolvent identities (see App.~\ref{app_resolvent}
for a more explicit derivation with slightly different notations) one can
show~\cite{Kl98,lelarge} that $G(z)$ obeys the following recursive 
distributional equation (RDE):
\begin{equation}
\label{rde}
G(z) \overset{\textnormal d}{=} 
\frac{1}{V-z-\sum_{i=1}^{k+1} J_i^2 \widetilde{G}_i(z)}\ ,
\hspace{1 cm}  
\widetilde{G}(z) \overset{\textnormal d}{=} 
\frac{1}{V-z-\sum_{i=1}^k J_i^2 \widetilde{G}_i(z)} \ .
\end{equation}
Here $\overset{\textnormal d}{=}$ denotes the equality in distribution
of random variables, $\widetilde{G}$ is the equivalent of $G$ on 
$\tT_k$ instead of $\T_k$, and the $\widetilde{G}_i$ (resp. $J_i,V$) are
i.i.d. copies of $\widetilde{G}$ (resp. $J$ and $V$ the distribution of
the off-diagonal and diagonal disorder). By definition $\E[G(z)]$
is the Stieltjes transform of $\rho$, hence the density of the absolutely
continuous part of $\rho$ (see the next subsection for a discussion of
the regularity properties of $\rho$) can be computed from the 
solution of the RDE as
\begin{equation}
\label{densite_resolvent} 
\rho_{\rm ac}(E) = 
\lim_{\eta \searrow 0}\frac{1}{\pi} \textnormal{Im} \mathbb{E} [G(E+i\eta)] \ .
\end{equation}
The equations (\ref{rde}) and (\ref{densite_resolvent}) can be solved 
numerically, and this is how we obtained the curves in 
Figs.~\ref{fig_numerical_simulation},\ref{fig_cusp} and \ref{fig_differents_k}.
We give more details on the numerical procedure in App.~\ref{appendix_middle}.

As mentioned in the previous section the density of states $\rho$
constructed here directly on the infinite regular tree coincides with
the large size limit of the empirical spectral measures for random
regular graphs. At variance with finite-dimensional models, the density
of states of the Bethe lattice cannot be obtained as the limit of growing
subgraphs of the infinite tree: the surface of the latter grows as fast as
their volume, which makes this construction ill-behaved.

We defined previously $u_n$ as the $n$'th moment of $\rho$. The expression
of $\rho$ that follows from 
Eqs.~(\ref{eq_def_spectral},\ref{eq_rho_def_infinite}) indicates that these 
moments 
can be computed on the infinite tree as $u_n = \E[\la 0 | H^n  | 0 \ra]$. 
We shall similarly define $\tu_n$ as $\E[\la 0 | H^n  | 0 \ra]$, where $H$
is now restricted on $\tT_k$ rooted at vertex 0, and denote these
moments $\tc_n$ (resp. $\td_n$) in the off-diagonal (resp. diagonal) case.

Let us note that in principle one could prove analytical results on the 
Lifshitz tail behavior of the density of states by characterizing directly
the solution of the RDE~(\ref{rde}). However it is a difficult task
to handle quantitatively this kind of equation, this is why we followed 
the indirect approach via the moments of the density of states in this
paper.

\subsection{Absolute continuity of the density of states}
\label{subsection_regularity}

\begin{figure}
\center
\includegraphics[width=8.2 cm]{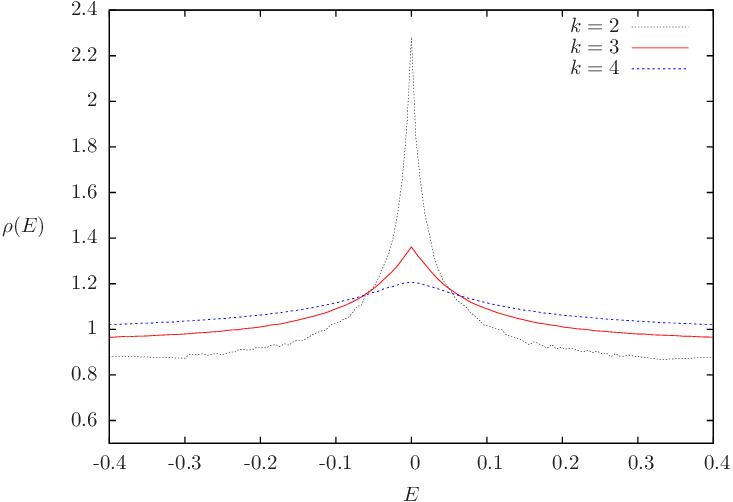}\hspace{0.8 cm} 
\includegraphics[width=8.2 cm]{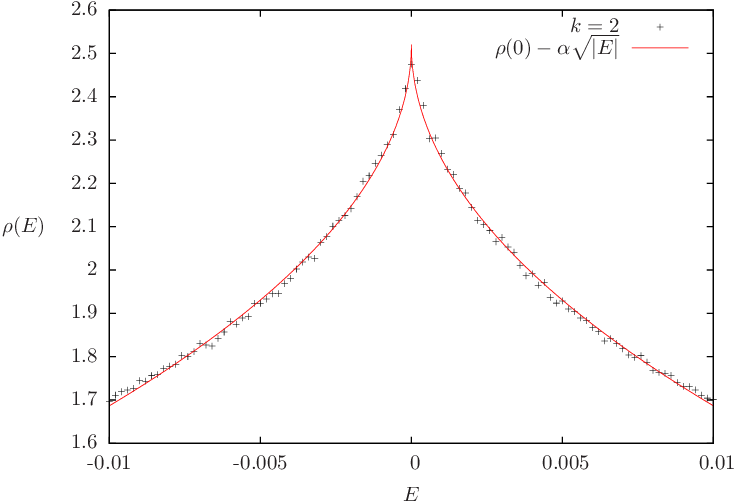}
\caption{Left: density of states around $E=0$ in the off-diagonal disorder 
case for 
$k=2,\;3,\;4$. To allow for comparison between these different values of
$k$ the couplings have been rescaled, and taken uniformly distributed
on $[-\sqrt{2/k},\sqrt{2/k}]$. Right: detail of the cusp around $E=0$ for 
$k=2$ and a fit to the analytic form derived heuristically in
Appendix \ref{appendix_middle}.}
\label{fig_cusp}
\end{figure}

As already underlined $\rho$ is in general a probability measure and as such
is not guaranteed to be absolutely continuous (i.e. to have a density).
However additional assumptions on the random variables $J_{ij}$ and $V_i$
besides (OD1-OD2) or (D1-D2)
are known to imply such a regularity of $\rho$ (this is why we denoted
$\rho(E)$ instead of $\rho_{\rm ac}(E)$ the density of states in 
Figs.~\ref{fig_numerical_simulation},\ref{fig_cusp} and 
\ref{fig_differents_k}). Let us first discuss
the diagonal disorder case. If $V$ has a bounded density, then $\rho$ is
absolutely continuous with a bounded density. This
is a well-known result for finite-dimensional models, usually called a
``Wegner estimate'' (see~\cite{Wegner} for the original work 
and~\cite{IDS,invitation,fabio} for mathematical presentations).
The core of the argument in finite dimension goes as follows: the number of 
eigenvalues of $H^L$ (the regularization of $H$ in a box of size $L$)
below a fixed threshold can only change by one
when the potential $V_i$ at one site is varied from its minimal to 
its maximal value. This is then shown to imply the absolute continuity
of the average of $N^L$, and finally of $\rho$ by taking the $L \to \infty$
limit. The argument can thus be readily extended to the Bethe lattice case
thanks to its regularization of finite size $N$ provided by the random 
regular graphs on $N$ vertices.
We stress that the absolute continuity of the density of states 
should not be mistaken with the nature (localized or extended) of the 
spectrum for a given realization of the disorder. 
 
Wegner's argument does not apply directly to the off-diagonal case, but under 
the additional assumption that $J$ has a Lipshitz continuous density the 
integrated density 
of states is locally Lipschitz continuous~\cite{klopp} (hence the density of 
states is almost everywhere defined and bounded), except possibly in the 
middle of the band (i.e. for $E=0$).
At this energy and in presence of off-diagonal disorder only, the density of 
states is known to diverge in one dimension~\cite{EgRi78}. 
The numerical estimates we obtained for the density of states 
of the Bethe lattice exhibit a weak non-analyticity around $E=0$ 
(see Fig.~\ref{fig_cusp}). A non-rigorous analysis presented 
in Appendix~\ref{appendix_middle} suggests the form
$\rho(E) \simeq \rho(0) - \alpha |E|^\frac{k-1}{2}$ when $E \rightarrow 0$, 
which is in very good agreement with the numerical results.

\subsection{From a probability density to its moments}
\label{sec_density_to_moments}

In the next subsection we shall present some bounds on the large $n$ behavior
of the moments $u_n$ of the density of states $\rho$, that reflects the 
behavior of $\rho$ around the edges of its support. To facilitate the 
intuitive understanding of the bounds we consider here the easy direction
of the connection between $\rho$ and $u_n$, that is we recall how the
behavior of a probability distribution at the border of its support
influences the growth of its moments.

Let us consider a probability density $\eta(E)$ supported on $[E_-,E_+]$
and its moments $u_n = \int_{E_-}^{E_+} E^n \eta(E) \dd E$.
In the large $n$ limit the dominant contribution to the integral arises 
from regions closer and closer to the edges $E_-$ and $E_+$, and thus the 
growth of the moments is controlled by the behavior of $\eta(E)$ 
close to the edge $E_-$ or $E_+$ which is largest in absolute value.
To simplify the discussion let us assume that $\eta$ is symmetric and denote
$E_0=E_+=-E_-$. Then one easily finds the following correspondences ($n$ is
implicitly even below):
\begin{itemize}
\item If $\eta(E) \sim (E_0-E)^\alpha$ ($\alpha >0$), then 
$\frac{1}{n} \ln u_n = \ln (E_0) - (\alpha+1) \frac{\ln n}{n} 
+ o\left(\frac{\ln n}{n}\right)$.
\item If $\eta(E) \sim e^{-\beta (E_0-E)^{-\alpha}}$ ($\alpha,\beta >0$), then 
$\frac{1}{n} \ln u_n = \ln(E_0) -c \, n^{-\frac{1}{\alpha+1}} 
+o(n^{-\frac{1}{\alpha+1}})$,
with $c$ a constant depending on $\alpha$ and $\beta$.
\item If $\eta(E) \sim e^{-\gamma e^{\beta(E_0-E)^{-\alpha}}}$ 
($\alpha,\beta ,\gamma>0$), then 
$\frac{1}{n} \ln u_n = \ln (E_0) - 
\frac{\beta^{1/\alpha}}{E_0} \frac{1}{(\ln n)^{1/\alpha}} 
+ o\left(\frac{1}{(\ln n)^{1/\alpha}}\right)$.
\end{itemize}
In all cases the value $E_0$ of the edge of the support controls the
dominant (exponential) growth of $u_n$, that is $E_0^n$,
while the behavior of $\eta$ in the neighborhood of $E_0$ yields
the subdominant corrections to $(\ln u_n)/n$. From one case to the next in 
these three examples $\eta(E)$ vanishes faster and faster as $E \to E_0$, 
and in consequence the corrections to $(\ln u_n)/n$ decay more and more 
slowly as $n \to \infty$. Note that in the last two cases the vanishing
of $\rho$ is fast enough for the integrated density of states $N(E)$ to
behave in an equivalent way at the leading order.

The Kesten-McKay distribution of Eq.~(\ref{eq_density_pure}) falls in the 
scope of the first case with $E_0 =2 \sqrt{k}$ and $\alpha =1/2$, so that
$\frac{1}{n} \ln{u_n} - \ln(2\sqrt{k}) \sim -\frac{3}{2} \frac{\ln{n}}{n}$ 
(which agrees with the exact expression for the moments obtained by 
McKay~\cite{mckay}, recalled in Sec.~\ref{sec_pure_case}).

As explained in the introduction one expects a doubly-exponential form
of the Lifshitz tail for the density of states on Bethe lattices, corresponding
to the third case above. This has
been established in a rigorous way in~\cite{muller,reinhold} for 
sub-critical bond percolation on the Bethe lattice, under the form
\begin{equation}
\lim_{\delta \to 0}\frac{\ln \ln |\ln(1-N(E_0-\delta))|}{\ln \delta} = 
- \frac{1}{2} \ ,
\end{equation}
where $N(E)$ is the integrated density of states. Its derivative $\rho(E)$
should behave in the same way (this has been proven in finite dimensions 
in~\cite{COGeKl10}), and
in consequence one should expect the correction terms in $(\ln u_n)/n$
to be of order $1/(\ln n)^2$. 
More precisely, the heuristic reasoning presented
in~\cite{KiHa85,BST10} suggests (for off-diagonal disorder)
$\rho(E) \sim (2\sqrt{k}-E)^{\left(\frac{k+1}{k}\right) k^{\pi k^{1/4}(2\sqrt{k}-E)^{-1/2}}}$,
hence one can apply the formula above with $E_0=2\sqrt{k}$, $\alpha=1/2$ and
$\beta=\pi k^{1/4} \ln k$ and make also a prediction for the coefficient of the 
correction term, $\frac{1}{n} \ln{c_n} - \ln(2\sqrt{k}) \sim 
-\frac{(\pi \ln k)^2}{2}\frac{1}{(\ln n)^2}$.
The prediction of~\cite{BST10} for diagonal disorder has the same form with
$E_0=2\sqrt{k}+W$, $\alpha=1/2$ and $\beta=\pi k^{1/4} \ln k$, which translates
into $\frac{1}{n} \ln{d_n} - \ln(2\sqrt{k}+W) \sim 
-\frac{(\pi \ln k)^2 \sqrt{k}}{2\sqrt{k}+W}\frac{1}{(\ln n)^2}$.

\subsection{Bounds on the moments and the integrated density of states}
\label{sec_res_bounds}

Let us announce here our main results, the proofs being postponed to 
sections \ref{sec_bounds_offd} and \ref{sec_bounds_d}. We recall that
the random variables defining the model are assumed to follow the
assumptions (OD1-OD2) in the off-diagonal disorder case (in particular
$J_{ij}$ has support $[-1,1]$) and (D1-D2) in the diagonal disorder case
($V_i$ is supported on $[0,W]$).

\medskip

\paragraph{Off-diagonal disorder}
\textit{For all $\epsilon,\; x>0$, there exists $n_0$ such that for all even $n > n_0$:}
 \begin{equation} 
-(1 + \epsilon) \frac{(\pi \ln k)^2}{2} \frac{1}{(\ln n)^2}
\leq \frac{1}{n} \ln c_n - \ln(2 \sqrt{k}) \leq 
- \frac{1}{(\ln n)^x}\frac{(\pi \ln k)^2}{2} \frac{1}{(\ln n)^2} \ .
\label{eq_statement_offd}
\end{equation}

\paragraph{Diagonal disorder}
\textit{For all $\epsilon,\; x>0$, there exists $n_0$ such that for all $n > n_0$:}
\begin{equation} 
- (1+\epsilon) \frac{(\pi \ln k)^2}{2}\frac{2 \sqrt{k} }{(2 \sqrt{k}+W)} 
\frac{1}{(\ln n)^2}  
\leq \frac{1}{n} \ln d_n - \ln(2 \sqrt{k} +W) \leq 
- \frac{1}{(\ln n)^x} \frac{(\pi \ln k)^2}{2}\frac{2 \sqrt{k} }{(2 \sqrt{k}+W)}
\frac{1}{(\ln n)^2} \ .
\label{eq_statement_d}
\end{equation}

Let us first precise the level of rigor of these results: the proofs
of the lowerbounds given in Sec.~\ref{sec_bounds_offd_lower} and 
Sec.~\ref{sec_bounds_d_lower} are mathematically rigorous. The upperbounds
rely on an explicit computation that we did not turn in a rigorous derivation,
but that we checked numerically with a very high accuracy, as discussed
more precisely in Sec.~\ref{sec_bounds_offd_upper} and 
Sec.~\ref{sec_bounds_d_upper}.

Note also that in the statement of the upperbounds the multiplicative constant
is actually irrelevant because of the condition $x>0$; we chose however to
write them in this suggestive form because we conjecture that the leading
term of the asymptotic expansion corresponds to $\epsilon=x=0$, i.e. that
\begin{equation}
\frac{1}{n} \ln c_n = \ln(2 \sqrt{k}) -
\frac{(\pi \ln k)^2}{2} \frac{1}{(\ln n)^2}
+o\left(\frac{1}{(\ln n)^2}\right) \ ,
\end{equation}
and a similar conjecture in the diagonal case, in full agreement with
the heuristic prediction on $\rho$ discussed above.

We now give the implication of these bounds on the moments
for the behavior of the density of states itself. 
The support of the latter being
bounded, $\rho$ is unambiguously determined by the knowledge of all its
moments~\cite{feller2}. The difficulty here is that we only have an 
asymptotic control on the moments, and that the very slow decay of their
corrections hampers the use of transfer (tauberian) theorems~\cite{feller2}. 
It is however possible to turn the above statements on the moments of $\rho$ 
into bounds on the integrated density of states $N(E) = \rho([-\infty,E])$:

\medskip

\paragraph{Off-diagonal disorder}
\textit{For all $\epsilon,\; x>0$, and $\delta$ small enough:}
\begin{equation}
\label{eq_statement_integrated_density_od} 
e^{-k^{\pi k^{1/4} \sqrt{\frac{  (1+ \epsilon)}{\delta}}}} \leq N(-2\sqrt{k}+ \delta) = 1-N(2 \sqrt{k}-\delta) \leq e^{-k^{\pi k^{1/4}\left({\frac{ 1}{\delta}}\right)^{\frac{1}{2+x}}}}
\end{equation}

\paragraph{Diagonal disorder}
\textit{For all $\epsilon,\; x>0$, and $\delta$ small enough:}
\begin{equation}
\label{eq_statement_integrated_density_d} 
e^{-k^{\pi k^{1/4} \sqrt{\frac{  (1+ \epsilon)}{\delta}}}} \leq 1-N(2 \sqrt{k}+W-\delta) \leq e^{-k^{\pi k^{1/4}\left({\frac{ 1}{\delta}}\right)^{\frac{1}{2+x}}}}
\end{equation}
The proofs of these two statements are deferred to Appendix~\ref{app_moments}. 
As mentionned in~\ref{sec_def}, the positivity of the support of $V$ in 
hypothesis (D1) simplifies the derivation of the bounds on the moments, but 
is not needed for (\ref{eq_statement_integrated_density_d}) to hold, this
last statement only requiring that $V$ is bounded and gives positive weight to 
the neighborhood of its upperbound $W$. Also note that as before the constants 
on the right-hand sides above are irrelevant due to the freedom in the choice 
of $x$, and are only here to suggest what we believe should be the asymptotic 
of $\ln | \ln(1-N(E_0-\delta))|$, with $E_0 = 2 \sqrt{k}$ in the 
off-diagonal disorder case and $2\sqrt{k}+W$ in the diagonal disorder case. 
In fact, a more modest and concise statement is:
\begin{equation}
\lim_{\delta \to 0}\frac{\ln \ln |\ln(1-N(E_0-\delta))|}{\ln \delta} = 
- \frac{1}{2} \ ,
\end{equation}
In particular this extends the result of~\cite{muller,reinhold} on the bond 
percolation model to the percolating phase.

Finally let us emphasize that in the course of the proof of the lower bounds
we shall obtain in a rigorous way that
\begin{equation}
\lim_{n \to \infty} \frac{1}{n} \ln c_n = \ln (2\sqrt{k}) \ , \qquad
\lim_{n \to \infty} \frac{1}{n} \ln d_n = \ln (2\sqrt{k} + W) \ ,
\end{equation}
with $n$ even in the first case. This is easily seen to imply 
(see~\cite{mckay} or App.~\ref{app_moments} for more details)
that the support of the density of states of the off-diagonal (resp. diagonal)
disorder model extends up to $2\sqrt{k}$ (resp. $2\sqrt{k} + W$), as
expected from the ergodicity arguments discussed above.

\section{Various expressions of the moments}
\label{sec_expressions}
\subsection{Walks on the tree}
\label{sec_walks_on_the_tree}

From the definition of the model on the (either finite or infinite) Bethe 
lattice it should be clear that the average moments $u_n$ (this notation
encompassing both the diagonal and off-diagonal disorder cases) can be written
as:
\begin{equation} 
u_n = \sum_{i_1, i_2, \dots, i_{n-1}}  \E [H_{0 i_1} H_{i_1 i_2} \dots H_{i_{n-1}0} ]
\ . 
\end{equation}
This is a sum over (lazy) walks of lengths $n$ on $\T_k$, 
starting and ending at its root $0$,
i.e. over sequences $0=i_0,i_1,\dots,i_{n-1},i_n=0$ of vertices of $\T_k$, such
that for all $j\in[0,n-1]$ either $i_j=i_{j+1}$ or $(i_j,i_{j+1})$ is an edge 
of $\T_k$. In the former case, that we shall call a self-bond step around vertex
$i_j$, the matrix element is $H_{i_j,i_{j+1}} = V_{i_j}$. In the latter case
$H_{i_j,i_{j+1}} = J_{i_j,i_{j+1}}$. As trees have no cycle each edge is visited
an even (possibly null) number of times during a closed walk. We denote
$\W_n$ the set of walks of length $n$ around the root $0$, and for a walk 
$\w= (i_1,\dots,i_{n-1})\in \W_n$
we define its weight $\pi$ as 
$\pi(\w) \equiv \E [H_{0 i_1} H_{i_1 i_2} \dots H_{i_{n-1}0} ]$, 
so that $u_n = \sum_{\w \in \W_n} \pi(\w)$.
A more explicit expression of $\pi$ is obtained by defining $n_e(\w)$
as half the number of times the edge $e\in \cE$ is crossed during the walk
$\w$, and $s_v(\w)$ the number of self-bond steps around vertex $v\in\cV$.
Indeed, as the disorder is given by i.i.d. random variables, 
\begin{equation} 
\pi(\w) = \prod_{e \in \cE} \E[J^{2 n_e(\w)}] \prod_{v\in \cV} \E[V^{s_v(\w)}] \ .
\end{equation}
In the off-diagonal disorder case only walks without self-bond steps have a 
non-zero weight; to avoid confusion we shall thus denote in this case
$\M_n$ the set of walks of length $n$ without self-bond steps.

It will be useful in the following (in particular in the proofs of the 
upperbounds in Sec.~\ref{sec_bounds_offd_upper} and \ref{sec_bounds_d_upper}) 
to refine the description of a walk, and to partition the set of walks 
$\W_n$ in various subsets.
We define the support $\sigma(\omega)$ of a walk $\omega$ as the set 
of edges of the tree visited at least once by $\omega$, i.e. 
$\sigma(\w)=\{ e \in \cE | n_e(\w) \ge 1 \}$, and the size of a support
as the number of edges it contains. A support of size $r$ is a subtree of
$\T_k$ containing the root and $r$ edges, i.e. a tree of $r$ edges where
the root has at most $k+1$ children, and all other vertices have at most
$k$ children. We shall denote $\S^r$ the set of the supports of size $r$.

The skeleton $\hsigma(\omega)$ is defined as the support of
$\omega$, supplemented by the numbers $\{n_e(\omega)\}_{e \in \sigma(\omega)}$.
More generally a skeleton $\hsigma$ is made of a support $\sigma$ and a set of
positive integers $\{n_e\}_{e \in \sigma}$; we call $2 \sum_{e \in \sigma} n_e$ 
the length of the skeleton $\hsigma$, and denote $\hS_n^r(\sigma)$ the set 
of skeletons of length $n$ based on a support $\sigma$ of size $r$.

We will also call self-support of a walk $\omega$ the set of vertices around
which at least one self-bond step is taken, i.e. the set
$\{ v\in \cV | s_v(\w) \ge 1\}$.

\begin{figure}[h]
\center
\includegraphics{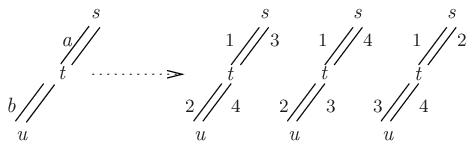}
\caption{A skeleton $\hsigma$ of size $2$ and length $8$, made of the support
$\sigma=\{a,b\}$ and the number of crossings of these two edges, $n_a = 2$ and
$n_b = 2$. The three vertices involved are denoted $\{s=0,t,u\}$, on the right
are drawn the three walks compatible with the skeleton (i.e. 
$\w=(s,t,u,t,s,t,u,t,s)$, $\w=(s,t,u,t,u,t,s,t,s)$ and 
$\w=(s,t,s,t,u,t,u,t,s)$), the numbers close to the edges being the order in 
which they are crossed from top to bottom.}
\label{fig_example_degeneracy}.
\end{figure}

The construction of a walk $\w \in \M_n$, i.e. of length $n$ without self-bond, 
amounts thus to the successive choices of:
\begin{itemize}
\item a support $\sigma$ of size $r \leq \frac{n}{2}$.
\item a positive integer $n_e$ for each edge $e$ of the support, such that 
$ 2 \sum_{e \in \sigma}  n_e =  n$. This completes the choice of the skeleton
of the walk.
\item an ordering of the visited edges, that is, a mapping $\varphi$ from 
$\{1, \dots, n\}$ to $\sigma$ compatible with the tree structure and the $n_e$.
In technical terms, one must have 
$\forall e \in \sigma, |\varphi^{-1}(\{e\})| = 2 n_e$ and 
$(\varphi(1), \dots, \varphi(n))$ must correspond to a walk on the vertices
covered by $\sigma$.
\end{itemize}
Note that there are in general several walks compatible with a given skeleton,
as explained on a simple example in Fig.~\ref{fig_example_degeneracy}; we 
shall come back to this issue in 
Sec.\ref{sec_bounds_offd_upper}-\ref{sec_bounds_offd_combinatorial_factor}.

The decomposition given above can be generalized to the case where self-bond
steps are allowed. Indeed, the definition of such a walk corresponds then
to choose:
\begin{itemize}
\item a number $s \in[0,n]$ of self-bond steps (such that $n-s$ is even).
\item a skeleton of length $n-s$.
\item an ordering of the edges of the support, satisfying the same 
constraints as given above in absence of self-bond steps. This
yields a walk $\ow \in \M_{n-s}$.
\item $s$ times $0\le t_1 \le t_2 \le \dots t_s \le n-s $, specifying that
self-bond steps have to be inserted after $t_1,t_2,\dots,t_s$ steps of 
$\ow$.
\end{itemize}

\subsection{A recursive computation}
\subsubsection{Recursion relation}
\label{sec_recursion_relation}

Taking advantage of the recursive structure of $\T_k$ one can 
set up a recursive computation of $u_n$. In this section we present the
equations and their intuitive interpretation, their formal derivation (which
is in essence a series expansion of the recursion (\ref{rde}) between the Green
functions) being deferred to App.~\ref{app_resolvent}. Similar recursive 
equations on the moments have been derived in~\cite{BaGo01,KhShVe04} for 
sparse random matrices, where the randomness lies in the degrees of the 
vertices.

Consider a closed walk of length $n$ starting at the root $0$ of $\T_k$,
call $s$ the number of self-bond steps made at $0$ and $p_i$ half
the number of times the edge between the root and its neighbor $i\in[1,k+1]$
is crossed. The walk will make a number $m_i$ of steps in each of the subtrees
$\T^{(i)}$, the copy of $\tT_k$ rooted at $i$, such that 
$s + \sum_{i} (m_i +2 p_i) =n$. One must however take into account the fact that
these $m_i$ steps are to be divided in $p_i$ closed walks on $\T^{(i)}$, 
separated by a visit to the root $0$. In each of these closed walks the 
disorder encountered in $\T^{(i)}$ is the same, hence the contribution of
the various closed walks on $\T^{(i)}$ are correlated. Summarizing these
observations one obtains the following expression of $u_n$,
\begin{equation} 
u_n = \sum_{\substack{
s,p_1,m_1,\dots,p_{k+1},m_{k+1} \ge 0\\
s+m_1+2 p_1 + \dots+ m_{k+1}+2 p_{k+1}=n}}
u(m_1, p_1) \dots u(m_{k+1},p_{k+1})  {s+p \choose s, p_1, \dots, p_{k+1}}  
\E[J^{2p_1}] \dots \E[J^{2 p_{k+1}}] \E[V^s] \ ,
\label{eq_recursion_vraie} 
\end{equation}
where in the multinomial coefficient we denoted $p=p_1+\dots+p_{k+1}$. This
coefficient counts the number of orderings of the steps made each time the
walk steps out the root, either for a self-bond step or towards one of the
$k+1$ neighbors. The quantity $u(m,p)$ is the average contribution of the
steps made in one of the subtrees $\T^{(i)}$, when $m$ steps are performed
in $\T^{(i)}$, divided in $p$ epochs separated by visits to the ancestor $0$.
A more explicit interpretation of $u(m,p)$ is the following: introduce an 
additional vertex $(-1)$ connected only to the root $0$ of $\tT_k$ with an edge
of weight $J_{-1,0}=1$, and define $u(m,p)$ as the weighted sum of the walks 
of length $m+2p$ on $\tT_k \cup{(-1)}$, that start and end
at 0 and that are constrained to visit exactly $p$ times the vertex $(-1)$.
Following the same reasoning as above one can compute the $u(m,p)$'s by
recursion, according to the rules
$ \hspace{0.3 cm} u(0,q) = 1, \hspace{0.5 cm} u(n,0) = \delta_{n,0}, $
\begin{equation} 
u(n,q) = \sum_{\substack{
s,p_1,m_1,\dots,p_k,m_k \ge 0\\
s+m_1+2 p_1 + \dots+ m_k+2 p_k=n}}
u(m_1, p_1) \dots u(m_k,p_k)  {q-1+s+p \choose q-1,s, p_1, \dots, p_k}  
\mathbb{E}[J^{2p_1}] \dots \mathbb{E}[J^{2 p_k}] \mathbb{E}[V^s] \ \ \
\text{for} \ \ q\ge 1 \ ,
\label{eq_recursion_cavite} 
\end{equation}
where again in the multinomial coefficient $p$ stands for $p_1+\dots+p_k$, and
this coefficient counts the number of orderings of the steps taken from the
root of $\T^{(i)}$ (there appears $q-1$ because the first passage of the walk 
on $i$ necessarily arrives from the ancestor).
Note that in particular $u(n,1)$ is equal to $\tu_n$, the moments of order
$n$ for the walks on $\tT_k$.

\subsubsection{Numerical evaluation}

One can check that Eq.~(\ref{eq_recursion_cavite}) does indeed provide a 
recursive scheme to compute all the $u(n,q)$~: the computation of these
values (and of the moment $u_n$ from Eq.~(\ref{eq_recursion_vraie})) 
at rank $n$ only requires the knowledge of $u(m,p)$ with $m<n$ and $2 p\le n$.
The number of terms to sum in order to obtain a new $u(n,q)$ grows as 
$O(n^{2k})$ ($O(n^{2k-1})$ in the off-diagonal case), and so the computation of 
$u_n$ requires a time that grows as $O(n^{2k+2})$ ($O(n^{2k+1})$ in the 
off-diagonal case). Though this is a polynomial time complexity the exponent is 
high (at least 5), and in practice only rather limited values of $n$ are 
accessible within a reasonable time on present computers. For illustration we 
present in Table~\ref{tab_results} numerical results up to $n=62$, in the 
off-diagonal disorder model with $k=2$, along with the corresponding values 
for the pure case.

\begin{table}[h]

\center
\begin{tabular}{|l|r|r|r|r|r|r|r|r|r|}

\hline
$n$ & 2&4 & 8& 16 & 24& 32&40&50& 62\\
 \hline
   pure & 2 & 8 &224& $3.66 \; 10^5$& $8.52 \; 10^8$ &$2.32 \; 10^{12}$& $6.88 \; 10^{15}$ & $1.63 \; 10^{20}$ & $3.12 \; 10^{25}$ \\

   disordered &$ 0.67 $& $1.06$ & $5.37$&375&$4.67 \; 10^4$&$7.93 \; 10^6$ &$1.66 \;10^9 $& $1.64 \; 10^{12}$ & $8.29 \; 10^{15}$ \\
   \hline
  
\end{tabular}

 \caption{Average value of the moments $\tilde{c}_n = u(n,1)$ on the 2-ary tree
$\tT_{k=2}$, in the off-diagonal disordered model with $J_{ij}$ uniformly random
on $[-1,1]$. The results of the pure case ($J_{ij}=1$) can be directly computed 
from Eq.~(\ref{eq_unq_pure_case}).}
\label{tab_results}
\end{table}

A lowerbound on $u_n$ can also be obtained numerically for larger values of 
$n$: as all terms in Eq.~(\ref{eq_recursion_cavite}) are positive, the
quantities $u'(n,q)$ obtained according to the recursion 
(\ref{eq_recursion_cavite}) in which all sums on $p_i$ are restricted to
$p_i \le p_{\rm max}$, a threshold fixed independently of $n$, are smaller than
the $u(n,q)$'s. The complexity of the computation of $u'(n,q)$ is reduced to
$O(n^{k+1})$ ($O(n^k)$ for the off-diagonal case); for $k=2$ we could reach
values of $n$ of the order of 200, using $p_{\rm max}=10$. The approximation 
coming from the finite value of $p_{\rm max}$ becomes worse and worse as $n$
grows larger. In any case, the values of $n$ reachable numerically remains
very far from the regime in which we expect the asymptotic scaling of the 
moments presented in Sec.~\ref{sec_res_bounds} to hold.

\subsection{Explicit expression}
\label{sec_explicit_expression}

By ``unfolding'' the recursive equations 
(\ref{eq_recursion_vraie},\ref{eq_recursion_cavite}), i.e. replacing the 
$u(n,q)$ in the r.h.s. by their recursive expressions, one obtains a formula
for the moment $u_n$ in which the summation over the number of visits of all
edges and self-bonds of the tree is made explicit. In order to simplify
the notations let us define a labelling on the edges and vertices of the 
tree by: the root has label $0$, the children of a vertex $v$ have indices 
$v_1, \dots, v_k$ (except the root whose neighbors are denoted $1,\dots,k+1$) 
and an edge that connects a vertex $v$ to its ancestor $v'$ has index $v$. 
With these definitions and the previously introduced notations, we obtain 
the following expression for $u_n$:
\begin{equation} 
u_n = \sum'_{\substack{\{ n_v \geq 0\}_{v \in \cV \setminus 0}\\
\{ s_v \geq 0\}_{v \in \cV}}}
{s_0+ n_1 + n_2 + \dots +  n_{k+1} \choose s_0, n_1, n_2, \dots,  n_{k+1}}
\E[V^{s_0}]
\prod_{v \in \cV \setminus 0} 
{ n_{v} - 1 + s_v + n_{v_1} + \dots +  n_{v_k}  \choose n_v-1, s_v, n_{v_1}, \dots, n_{v_k}}
\E[J^{2 n_v}] \E[V^{s_v}] \ ,
\label{eq_moment_deplie} 
\end{equation}
where the prime on the sum denotes the constraint 
$s_0 + \underset{v \in \cV \setminus 0}{\sum} (2 n_v + s_v) = n$ that has
to be verified by the summands. We also used the convention 
${\dots \choose -1, s, n_1, \dots, n_k} = \delta_{s,0}\delta_{n_1,0} \dots \delta_{n_k,0}$, which arises from the boundary condition $u(n,0)=\delta_{n,0}$.
More explicitly, this convention for the multinomial coefficient enforces the
connected character of the set of non-zero elements $n_v$, which in consequence
form a valid skeleton for the walks (incidentally this also implies that the 
number of non-zero terms in the sum is finite for any finite $n$). 
This expression can be thus interpreted
as a sum over skeletons and number of self-bond steps of a walk, the product
of the multinomial coefficients counting the number of walks compatible with
such values of $\{n_e\}_{e \in \cE}$, $\{s_v\}_{v \in \cV}$, that arises from
the freedom of choice in the order the steps around each vertex are taken.
Let us remark finally that this expression can easily be generalized to an 
arbitrary tree, at the price of slightly more cumbersome notations.

\subsection{The moments of the pure case}
\label{sec_pure_case}

As mentioned above  the density of states is given in the pure case 
($J_{ij}=1$, $V_i=0$) by the Kesten-McKay law whose density was recalled in 
Eq.~(\ref{eq_density_pure}). As a matter of fact this measure was determined
by McKay~\cite{mckay} via the computation of its moments. Let us briefly
recall this derivation, that we shall extend in 
Sec.~\ref{sec_bounds_offd_lower} and \ref{sec_bounds_d_lower} to obtain the
lowerbounds on the moments in the disordered case. We denote $c_n^0$ and
$\tc_n^0$ the values of the moments on $\T_k$ and $\tT_k$ respectively
for this pure case, as well as $u^0(n,q)$ for the solution of 
Eq.~(\ref{eq_recursion_cavite}).

\begin{figure}[h]
\center
\includegraphics[width=12cm]{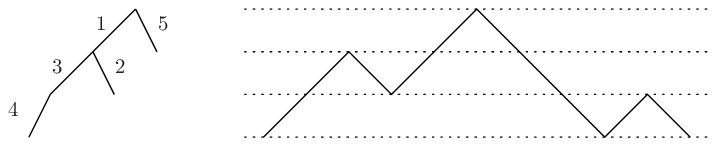}
\caption{A walk on the $2$-ary tree $\tT_2$ and the corresponding Dyck path.
In the left figure the numbers give the order in which the edges are crossed
from top to bottom.}
\label{walk_dyck}
\end{figure}

Consider a closed walk on $\tT_k$ of length $n$, 
$\w=(0=i_0,i_1,\dots,i_{n-1},i_n=0)$, and call depth of a vertex the distance
that separates it from the root. Then the sequence 
$(0=d_0,d_1,\dots,d_{n-1},d_n=0)$ of the depths of the vertices visited by
$\w$ forms a Dyck path, i.e. they are non-negative integers, $|d_{i+1}-d_i|=1$,
with 0 as final and initial values. This correspondence is illustrated in 
Fig.~\ref{walk_dyck}.

The computation of the number of Dyck paths of length $n$ is a classical
exercise in combinatorics~\cite{feller1,flajolet}, which yields the Catalan 
number ${n \choose n/2}\frac{1}{n/2 +1}$ for $n$ even, 0 otherwise. Coming
back to the computation of $\tc_n$, the important point to realize is that
there are exactly $k^{n/2}$ walks on $\tT_k$ that give the same Dyck path of
length $n$: at every of the $n/2$ steps taken away from the root there are
$k$ possible children to choose from. Hence we obtain in the pure case
\begin{equation}
u^0(n,1)=\tc_n^0 = {n \choose n/2}\frac{1}{n/2 +1} k^{n/2}
\label{eq_tcn_0}
\end{equation}
for $n$ even, zero otherwise.

A similar computation can also be done on $\T_k$; the depth projection of
a closed walk of length $n$ on $\T_k$ still yields a Dyck path of length $n$.
However the number of distinct walks corresponding to the same Dyck paths
depends on the number of its returns to the origin. Indeed there are $k+1$
choices for a step out of the root, and only $k$ for steps taken away from the
root from another vertex. The number of Dyck paths of length $n$ with $i$ 
returns to the origin can also be enumerated~\cite{feller1}, the result being
${n-i \choose n/2}\frac{i}{n-i}$. This yields the moment
\begin{equation}
c_n^0 = \sum_{i=1}^{n/2} {n-i \choose n/2}\frac{i}{n-i} (k+1)^i k^{n/2 -i} \ ,
\end{equation}
which can be checked to be the $n$-th moment (for $n$ even) of the 
Kesten-McKay distribution of Eq.~(\ref{eq_density_pure}). A crude estimation
of the asymptotic behavior of $c_n^0$ shows that 
$\lim \frac{1}{n} \ln c_n^0 = \ln(2 \sqrt{k})$, 
which reflects the fact that the 
support of the distribution in Eq.~(\ref{eq_density_pure}) is $2 \sqrt{k}$
(recall the discussion in Sec.~\ref{sec_density_to_moments}). The same
asymptotic behavior is readily found for $\tc_n^0$ from Eq.~(\ref{eq_tcn_0}).

In the pure case one can actually find the solution of the recursive 
equation~(\ref{eq_recursion_cavite}) by similar considerations on the
depth projections of closed walks,
\begin{equation}
u^0(n,q) = {n+q \choose n/2 +q} \frac{q}{n+q} k^{n/2}
\label{eq_unq_pure_case} 
\end{equation}
for $n$ even, zero otherwise.
We show in Appendix~\ref{app_pure_case} an explicit verification of this
statement.

Let us make a final series of remarks before closing this section:
\begin{itemize}
\item the projection from a walk on a tree to the Dyck path representing
the distance from the root to the walker is a powerful tool in the pure case
because it is easy to count the number of walks corresponding to a given Dyck
path. Unfortunately it is much harder to compute the total weight of walks 
sharing a Dyck path in the (off-diagonal) disorder case: the disorder is
kept fixed along the walk, this induces correlations between the choices
of the branch followed by the walk which are lost in the projection on its
depth.

\item one can try to apply a saddle-point estimate  to the recursion 
equation (\ref{eq_recursion_cavite}): writing 
$u(n,q=nx) \simeq e^{nf(x)}$  gives a functional equation on $f(x)$ 
whose solution is found to be, independently of the disorder, the one 
of the pure case $f(x) = \ln k + (2+x)\ln (2+x)-(1+x)\ln (1+x)$. However, 
we were not able to get the non-exponential corrections to this formula in 
the disordered case, as the difference in (\ref{eq_recursion_cavite}) between 
the pure and the disordered case is only in polynomials factors, whereas
according to \ref{sec_res_bounds} we expect a slower dominant change in 
$u(n,q)$. Another difficulty is that the corrections to $u(n,q)$ do not 
depend much on the particular shape of disorder, whereas 
(\ref{eq_recursion_cavite}) seems to. An alternative approach that we
did not pursue would be to seek a specific distribution of the disorder such
that its moments make Eq.~(\ref{eq_recursion_cavite}) solvable with a closed
combinatorial form for $u(n,q)$.

\item the asymptotic behavior of the ``shifted pure case'' where $J_{ij}=1$, 
$V_i=W$ a fixed constant, for which the moments are denoted $d_n^0$, can
be obtained either by shifting the density of Eq.~(\ref{eq_density_pure})
to $\rho_{\rm KM}(E-W)$ or by depth projections leading to Motzkin paths
(see App.~\ref{app_lower_bound} for a definition) with a weight $W$
assigned to self-bond (horizontal) steps. This yields 
$\lim \frac{1}{n} \ln d_n^0 = \ln(2 \sqrt{k} + W)$, a result which shall be
used for comparison in the discussion of the diagonal disorder model.
\end{itemize}

\section{Proofs for the off-diagonal disorder model}
\label{sec_bounds_offd}

This section contains the proofs of the bounds stated in 
Eq.~(\ref{eq_statement_offd}) 
for the off-diagonal disorder case, i.e. for vertex random variables $V_i$ all
vanishing, and random edge couplings $J_{ij}$ drawn in $[-1,1]$ with a
probability distribution satisfying the assumptions (OD1-OD2) of 
Sec.~\ref{sec_def}. We denote $c_n$ the moments of the density of states,
and assume implicitly below that $n$ is even, as all odd moments obviously
vanish in this case, the length of a closed walk on a tree being by definition
even.

\subsection{Lower bound}
\label{sec_bounds_offd_lower}

Let us first explain the strategy underlying the proof of the lowerbound,
reminiscent of the procedure for finite-dimensional models sketched in the
introduction. We shall indeed lowerbound the moments $c_n$ of the density of
states by restricting the contributions of walks to those confined to a 
small volume around the root (at a distance smaller than $h$), under the 
condition that the disorder takes a large value in this volume (i.e. all
$|J_{ij}|$ shall be greater than $J_0$). An optimal compromise will then be 
found between these two restrictions: taking $h$ too small reduces too 
drastically the possible contributions to $c_n$, while the probability of 
having large disorder falls off quickly if $h$ is taken too large. We now
turn this reasoning into an explicit derivation.

We recall that $c_n = \sum_{\omega \in \M_n} \pi(\omega)$, where
$\M_n$ denotes the set of closed walks (with no self-bond steps) of length 
$n$ starting at the root of $\T_k$. The weight of a walk $\pi(\omega)$ is
given by $\E[\prod_e J_e^{2 n_e(\omega)} ]$, where $e$ runs over the edges of
$\T_k$ and $n_e(\omega)$ is a non-negative integer equal to half the number
of times the walk crosses the edge $e$. As all weights $\pi(\omega)$ are 
positive, $c_n$ can be lowerbounded by restricting the sum to $\tM_n$, the
walks on $\tT_k$, and further to $\tM_{n,h}$, the walks on $\tT_k$ which go
at a distance at most $h$ from the root:
\begin{equation} 
c_n \geq \tc_n = \sum_{\omega \in \tM_n} \pi(\omega) \geq 
\sum_{\omega \in \tM_{n,h}} \pi(\omega) \ .
\end{equation}
Let us denote $\cE_h$ the set of edges with endpoints at distance at most
$h$ from the root of $\tT_k$, $M$ the event: 
$\{|J_e|\ge J_0 \ \forall e \in \cE_h\}$, where $J_0$ is an arbitrary threshold
with $0<J_0<1$, and $\nM$ the complementary event. 
For an arbitrary walk $\omega$ in $\tM_{n,h}$ we have
\begin{equation}
\begin{split}
\pi(\omega) & = \E\left[\left. \prod_{e \in \cE_h} J_e^{2 n_e(\omega)} \right| M \right] 
\P[M] + \E\left[\left. \prod_{e \in \cE_h} J_e^{2 n_e(\omega)} \right| \nM \right] 
\P[\nM] \\
&\ge \E\left[\left. \prod_{e \in \cE_h} J_e^{2 n_e(\omega)} \right| M \right] 
\P[M] \\
& \ge J_0^n \, \P[|J|\ge J_0]^{|\cE_h|} \ ,
\end{split}
\label{eq_offd_bound_pi}
\end{equation}
where in the last step we have used the independence of the $J_e$ on different
edges and the fact that $\sum_e 2 n_e(\omega) = n$. This lowerbound being 
independent of $\omega$, it is now enough to control the number $|\tM_{n,h}|$ 
of walks on $\tT_k$ which remain at a distance smaller or equal to $h$ 
from the root. Using the projection from a walk to its depth explained in 
Sec.~\ref{sec_pure_case},
one realizes that $|\tM_{n,h}| = k^{n/2} m_{n,h}$, where $m_{n,h}$ is the
number of Dyck paths of length $n$ and of height at most $h$.
These paths can be enumerated with standard combinatorial 
techniques~\cite{krattenthaler, flajolet}; in particular we show in 
App.~\ref{app_lower_bound} (along with a more precise estimate of its 
asymptotic behavior) that $m_{n,h}$ obeys the inequality
\begin{equation}
m_{n,h} \ge \left(\frac{2}{h+2}\right)^3 
\left(2 \cos\left(\frac{\pi}{h+2} \right) \right)^n \ ,
\end{equation}
valid for all values of $h$ and of (even) $n$. We can thus write
\begin{equation}
c_n \ge J_0^n \, \P[|J|\ge J_0]^{|\cE_h|} \, \left(\frac{2}{h+2}\right)^3 
\left(2 \sqrt{k} \cos\left(\frac{\pi}{h+2} \right) \right)^n \ ,
\end{equation}
or equivalently after taking logarithm, subtracting the leading order
of the pure case and using the fact that 
$|\cE_h|=\frac{k^{h+1}-k}{k-1} \le \frac{k^{h+1}}{k-1}$:
\begin{equation}
\frac{1}{n} \ln c_n  - \ln(2 \sqrt{k}) \ge \ln(J_0) +
\frac{k^{h+1}}{n(k-1)} \ln \P[|J|\ge J_0] + 
\frac{3}{n} \ln \left(\frac{2}{h+2}\right) +
\ln \cos\left(\frac{\pi}{h+2} \right) \ .
\label{eq_lower_bound_off_diag}
\end{equation}

From this equation we shall first show that 
$\underset{n \to \infty}{\lim} \frac{1}{n} \ln c_n = \ln(2 \sqrt{k})$. 
Fixing $h$ and $J_0$
and letting $n \to \infty$ in the inequality above yields
\begin{equation}
\underset{n \to \infty}{\liminf} \frac{1}{n} \ln c_n  - \ln(2 \sqrt{k}) 
\ge \ln(J_0) + \ln \cos\left(\frac{\pi}{h+2} \right) \ .
\end{equation}
Sending now $J_0$ to 1 and $h$ to infinity we obtain 
$\liminf \frac{1}{n} \ln c_n  \ge \ln(2 \sqrt{k})$. On the other hand we 
obviously have $c_n \le c_n^0$, the value in the pure case where all 
$J_{ij}=1$, and
the results of McKay~\cite{mckay} recalled in Sec.~\ref{sec_pure_case} implies
that $\frac{1}{n} \ln c_n^0 \to \ln(2 \sqrt{k})$. From these matching lower and
upper bounds it follows that 
$\lim_{n \to \infty} \frac{1}{n} \ln c_n = \ln(2 \sqrt{k})$, i.e. that the edge
of the density of states is the same as in the pure case, as expected for
the ergodicity reasons discussed in Sec.~\ref{sec_def_model_infinite}.
Note that the only assumption
on the distribution of $J$ which was necessary here is the existence of some
$J_{\rm min}\ge 0$ such that $\P[|J|\ge J_0] >0$ for all
$J_0$ with $J_{\rm min} \le J_0 <1$.

The more precise result stated in Eq.~(\ref{eq_statement_offd}) is obtained
by taking $h \to \infty$ and $J_0 \to 1$ in an $n$-dependent way. By inspection
of the last term in the r.h.s. of Eq.~(\ref{eq_lower_bound_off_diag}) one 
realizes that the best bound will be achieved by taking $h$ as large as 
possible. The limitation on the possible range of $h$ comes instead from the 
second term, in which $k^{h+1}$ has to be small compared to $n$.
We shall in consequence take $h=\lfloor\alpha \ln n \rfloor$ 
with $\alpha< 1/\ln k$ and 
$J_0=1-\frac{1}{(\ln n)^{2+y}}$ where $y>0$ is an arbitrary positive constant.
In the limit $n \to \infty$ the r.h.s. of Eq.~(\ref{eq_lower_bound_off_diag})
becomes
\begin{equation}
-\frac{\pi^2}{2 \alpha^2} \frac{1}{(\ln n)^2} + \frac{k}{k-1} 
\frac{1}{n^{1-\alpha \ln k}} \ln \P\left[|J|\ge 1 - \frac{1}{(\ln n)^{2+y}}\right]
+ O \left( \frac{1}{(\ln n)^{2+y}} \right) \ .
\end{equation}
For all values of $\alpha< 1/\ln k$ and $y>0$ the second term is, thanks to
the assumption (OD2) on the random variable $J$, negligible with respect to
the first, which proves the statement of the lowerbound in
Eq.~(\ref{eq_statement_offd}). 

If the assumption (OD2) were violated the
dominant effect at the origin of the reduction of the density of states
near the edge would be the extremely low probability for a single $|J_{ij}|$
to be close to 1, not the collective effect of all $|J_{ij}|$ in a given volume 
being large.
In that case the form of the lowerbound in Eq.~(\ref{eq_statement_offd}) 
would have to be modified and would depend on the precise form of the
density of $J$ around 1. The transition from the collective to the single edge
dominated regime of the Lifshitz tail phenomenon, depending on the form of
the edge strength distribution near its edge, has been studied in the 
finite-dimensional case in~\cite{PaFi}. The methods developed in this paper,
in particular for the upperbound proofs, do not seem to us well adapted to
the single edge dominated regime, which shall not been discussed further here.

\subsection{Upper bound}
\label{sec_bounds_offd_upper}

The derivation of the upperbound in Eq.~(\ref{eq_statement_offd}) will proceed
by a decomposition of the sum over the walks $\w \in \M_n$ according to the
size $|\sigma(\w)|$ of their support, i.e. the number of edges of the tree 
visited at least once by $\w$ (recall the definitions given in 
Sec.~\ref{sec_walks_on_the_tree}). Indeed, the weight of a walk can be 
upperbounded as follows:
\begin{equation}
\pi(\w) = \prod_{e \in \sigma(\w)}\E[J^{2 n_e(\omega)}] 
\le \prod_{e \in \sigma(\w)}\E[J^2] = \E[J^2]^{|\sigma(\w)|} \ ,
\end{equation}
where we have used the facts that $|J| \le 1$ and that (by definition of the
support) $n_e(\w) \ge 1$ for $e \in \sigma(\w)$. By our assumption (OD1)
the variance $\E [J^2]$ is strictly smaller than 1, hence the
weight of a walk is exponentially small in the size of its support. The 
strategy of the proof corresponds then to choose in an optimal way a 
($n$-dependent) threshold on the support's size, use the bound above for
walks with larger supports, and upperbound the number of walks with smaller
supports.

Let $r_{\rm m}(n)$ be an increasing (integer) function of $n$ to be fixed later,
such that $\lim r_{\rm m}(n) = \infty$ and 
$\lim \frac{r_{\rm m}(n)}{n} = 0$ as $n \to \infty$.
We define $\M_n^\le \subset \M_n$ as 
the set of walks 
$\omega$ whose support has size at most $r_{\rm m}(n)$. Using the obvious 
fact that $\pi(\w) \le 1$ for any walk, we obtain
\begin{equation}
c_n \le \E[J^2]^{r_{\rm m}(n)} |\M_n \setminus \M_n^\le| + |\M_n^\le| 
\le \E[J^2]^{r_{\rm m}(n)} |\M_n | + |\M_n^\le| \ .
\label{eq_cn_upper}
\end{equation}
According to the discussion of Sec.~\ref{sec_walks_on_the_tree}, 
the number of walks in $\M_n^\le$ can be written as
\begin{equation}
|\M_n^\le| = \sum_{r=1}^{r_{\rm m}(n)} \sum_{\sigma \in \S^r}
\sum_{\hsigma \in \hS_n^r(\sigma)} \eta(\hsigma) \ ,
\label{eq_Wnle_1}
\end{equation}
where $\S^r$ is the set of supports of size $r$, $\hS_n^r(\sigma)$ the set 
of skeletons of length $n$ based on a support $\sigma$ of size $r$, and
$\eta(\hsigma)$ the number of walks compatible with such a skeleton.
The rest of the proof will rely on the following statement about the maximal 
value $\kappa(n)$ of the combinatorial factor $\eta(\hsigma)$:

\textit{Let $\kappa(n) = \max \{\eta(\hsigma) | r \in [1,n/2], \sigma \in \S^r, \hsigma \in \hS_n^r(\sigma) \} $. Then for all $y>0$ and
$\alpha <(\pi \ln k)^2/2$, it holds for $n$ large enough that:} 
\begin{equation} 
\label{eq_lemme_kappa} 
(2 \sqrt{k})^n e^{- \frac{n}{(\ln{n})^{2-y}}} \leq \kappa(n) 
\leq  (2 \sqrt{k})^n e^{-\alpha \frac{n}{(\ln{n})^2}}  \ .
\end{equation}

It should be emphasized that the total number of walks of length $n$ is, at 
the leading exponential order, $(2 \sqrt{k})^n$, so that the previous 
statement shows that an unique skeleton $\hsigma$ with very large 
combinatorial factor $\eta(\hsigma)$ contributes to the leading 
exponential order to the sum over all walks. In the next subsection
we explain how to obtain these bounds on $\kappa(n)$; 
unfortunately we were not able to derive the upperbound in 
Eq.~(\ref{eq_lemme_kappa}) in a fully rigorous way, yet it is the result
of an analytical computation that we also checked numerically
with high precision. Hence we continue here the derivation
assuming that Eq.~(\ref{eq_lemme_kappa}) holds true.

Going back to Eq.~(\ref{eq_Wnle_1}) and using the definition of $\kappa(n)$
yields
\begin{equation}
|\M_n^\le| \le \kappa(n) 
\sum_{r=1}^{r_{\rm m}(n)} 
\sum_{\sigma \in \S^r} |\hS_n^r(\sigma)| \ .
\label{eq_Wnle_2}
\end{equation}
It is easy to see that $|\hS_n^r(\sigma)|= {n/2 -1 \choose r-1}$ 
independently of $\sigma$: this is the number of ways to choose $r$
positive integers (the $\{n_e\}_{e \in \sigma}$) which sum to $n/2$.
The numbers $|\S^r|$ of supports of size $r$ are clearly increasing with $r$.
Moreover $r_{\rm m}(n)<n/2$ for $n$ large enough, hence the terms of the sum in
Eq.~(\ref{eq_Wnle_2}) can be upperbounded by their value in $r_{\rm m}(n)$:
\begin{equation}
|\M_n^\le| \le 
\kappa(n) r_{\rm m}(n) {n/2 -1 \choose r_{\rm m}(n)-1} |\S^{r_{\rm m}(n)} | \ .
\label{eq_Wnle_3}
\end{equation}
The numbers $|\S^r|$ of supports of size $r$ can be enumerated via
their generating function $F(x)=\sum_r |\S^r| x^r$. A support of size $r$ being 
a subtree of $\T_k$ of $r$ edges, it is composed of a number $j \in [0,k+1]$ of
edges around the root, along with $j$ subtrees of the copy of $\tT_k$ rooted
at the neighbors of the root. This implies that $F(x)=(1+x\tF(x))^{k+1}$,
where $\tF$ is the equivalent of $F$ for $\tT_k$, solution of 
$\tF(x) = (1+x \tF(x))^k$. The asymptotic behavior of $|\S^r|$ for large
$r$ can be inferred from the analysis of the singularities of the generating
function (see Theorem VII.3 in~\cite{flajolet}). An elementary study of
the reciprocal function $x(\tF)=(\tF^{1/k}-1)/\tF$ shows that it has a 
maximum equal to $1/\gamma_k$, with $\gamma_k = \frac{k^k}{(k-1)^{k-1}}$, hence
$\tF(x)$ has a square-root singularity in $x=1/\gamma_k$. As $F(x)$ is known
explicitly in terms of $\tF$ one obtains easily
\begin{equation}
F(x) = A_k - B_k \sqrt{1-\gamma_k x} + O(1-\gamma_k x) 
\quad \text{as} \ \ x \to 1/\gamma_k \ ,
\end{equation}
where $A_k$ and $B_k$ are two positive constants whose precise expression
will not be useful in the following. This singularity is then translated
in terms of the $|\S^r|$ as~\cite{flajolet}:
\begin{equation}
|\S^r| = \frac{B_k}{2 \sqrt{\pi r^3}} \gamma_k^r (1+O(r^{-1})) \ . 
\end{equation}

We now choose an arbitrary $x>0$, and set 
$r_{\rm m}(n)= \lfloor \frac{n}{(\ln n)^{2+x}} \rfloor$. With this choice
for the maximal size of the supports in $\M_n^\le$, one has
\begin{equation}
\frac{1}{n} \ln\left[
r_{\rm m}(n) {n/2 -1 \choose r_{\rm m}(n)-1} |\S^{r_{\rm m}(n)} | 
\right] = O\left( \frac{\ln \ln n}{(\ln n)^{2+x}}  \right) \ .
\label{eq_others}
\end{equation}
Hence the two leading orders of the inequality in (\ref{eq_Wnle_3}) arises
from $\kappa(n)$. Thanks to the freedom of choice of $\alpha$ in 
Eq.~(\ref{eq_lemme_kappa}) we conclude that for all $\beta < (\pi \ln k)^2/2$ 
and $n$ large enough: 
\begin{equation}
|\M_n^\le| \leq (2 \sqrt{k})^n e^{-\beta \frac{n}{(\ln n)^{2}}} \ .
\label{eq_domination_grands_squelettes} 
\end{equation}
Dividing Eq.~(\ref{eq_cn_upper}) by $c_n^0= |\M_n|$, the moment of the 
pure-case, one obtains for $n$ large enough,
\begin{equation}
\frac{c_n}{c_n^0} \le e^{\lfloor \frac{n}{(\ln n)^{2+x}} \rfloor \ln(\E[J^2]) } 
+ e^{-\beta \frac{n}{(\ln{n})^{2}}} \ ,
\label{eq_cn_upper2}
\end{equation}
using the fact that the sub-exponential corrections to $c_n^0$ are negligible
here (recall that 
$\frac{1}{n} \ln(c_n^0) = \ln(2 \sqrt{k})+ O(\frac{\ln n}{n}) $). In this
last inequality the first term is the dominating one, and leads to 
\begin{equation}
\frac{1}{n} \ln (c_n) \le \ln(2 \sqrt{k}) 
- \frac{1}{(\ln n)^{2+x}} \ln (\E[J^2]^{-1}) \ .
\label{eq_upper_bound_off_diag_proof}
\end{equation}
As $x$ is constrained to be positive the (positive) constant 
$\ln (\E[J^2]^{-1})$ is irrelevant for $n$ large enough; this yields
the upperbound stated in Eq.~(\ref{eq_statement_offd}).

Let us finally comment on the possible improvement of this upperbound.
With a better estimate of $|\M_n \setminus \M_n^\le|$ one should try to
make the threshold function $r_{\rm m}(n)$ grow faster with $n$.
The limiting factor would then
be the allowed range of $\beta$ in the second term of Eq.~(\ref{eq_cn_upper2}),
$\beta < (\pi \ln k)^2/2$; this would actually be the best one can hope for,
as a greater value of $\beta$ would contradict
the lowerbound in Eq.~(\ref{eq_statement_offd}).

\subsection{Combinatorial factor}
\label{sec_bounds_offd_combinatorial_factor}

We shall now present our arguments in favor of the right inequality in 
(\ref{eq_lemme_kappa}). We recall that the expression of $\eta(\hsigma)$,
i.e. the number of walks compatible with the skeleton $\hsigma$, is given
by the product of the multinomial coefficients in Eq.~(\ref{eq_moment_deplie}).
The maximal value of $\eta$ can thus be written as
\begin{equation} 
\kappa(n) = \max_{r\in [1,n/2]} \max_{\sigma \in \S^r} 
\max_{\substack{\{n_e > 0\}_{e \in \sigma} \\ 2 \sum n_e = n}}
\left[ {n_1 + n_2 + \dots +  n_{k+1} \choose n_1, n_2, \dots,  n_{k+1}}
\prod_{v \in \sigma \setminus 0} 
{ n_{v} - 1 + n_{v_1} + \dots +  n_{v_k}  \choose n_v-1, n_{v_1}, \dots, n_{v_k}}
\right] \ ,
\label{eq_kappa}
\end{equation}
where we follow the labelling of $\T_k$ introduced in 
Sec.~\ref{sec_explicit_expression}, giving to an edge the index of its 
endpoint vertex most distant from the root. We also set by convention 
$n_e=0$ if $e\notin \sigma$.

An upperbound on $\kappa(n)$ is obtained by relaxing the constraint on the
possible values of the variables $\{n_e\}$. We shall thus extend their domain
to non-negative reals, denoted ${x_e}$ to avoid confusion, using the natural
extension $\Gamma(x+1)=x!$ from integer $x$ to arbitrary reals. We also let
the set of edges with positive values of $x_e$ to be an arbitrary subset of
$\cE_p$, which contains the edges with endpoint at distance at most $p$ from
the root. For the ease of notation we introduce $\varkappa(n)=\ln \kappa(n)$,
which is thus upperbounded as
\begin{equation}
\begin{split}
\varkappa(n) \le \max_{p \in [1,n/2]} 
\sup_{\substack{\{x_e \ge 1 \}_{e \in \cE_p} \\ 2 \sum x_e = n}}
& \bigg[ \ln\Gamma(1+x_1+x_2+\dots+x_{k+1}) - \ln\Gamma(1+x_1)-\dots - 
\ln\Gamma(1+x_{k+1}) \\ &
+ \sum_{v\in \cE_{p-1} \setminus 0}[ \ln\Gamma(x_v + x_{v_1} + \dots + x_{v_k} )
- \ln\Gamma(x_v) - \ln\Gamma(1+x_{v_1}) - \dots - \ln\Gamma(1+x_{v_k})] \bigg]\ .
\end{split}
\label{eq_varkappa} 
\end{equation}
We expect this upperbound to become
tight at the leading order when $n\to \infty$: the integer values of $n_e$ 
achieving the maximum in Eq.~(\ref{eq_kappa}) will become large as well, 
hence relaxing them to be real numbers should have a minor effect.

\begin{figure}
\center
\includegraphics[width=7cm]{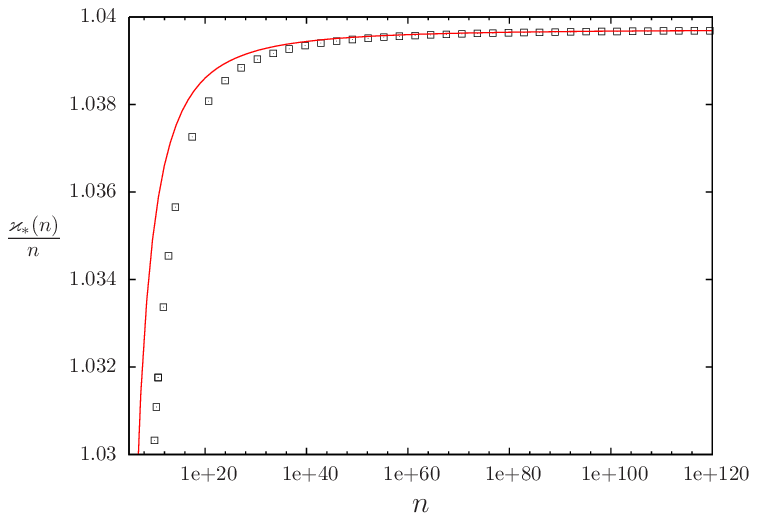}
\caption{
Plot of the function $\varkappa_*(n)$ defined in 
Eq.~(\ref{eq_def_varkappastar}), the symbols are the result of the numerical
evaluation of $\varkappa_*(n)/n$, the line is the conjecture 
(\ref{eq_conjecture_varkappastar}). These results are for $k=2$.}
\label{fig_varkappa}
\end{figure}

\begin{figure}
\center
\includegraphics[width=7cm]{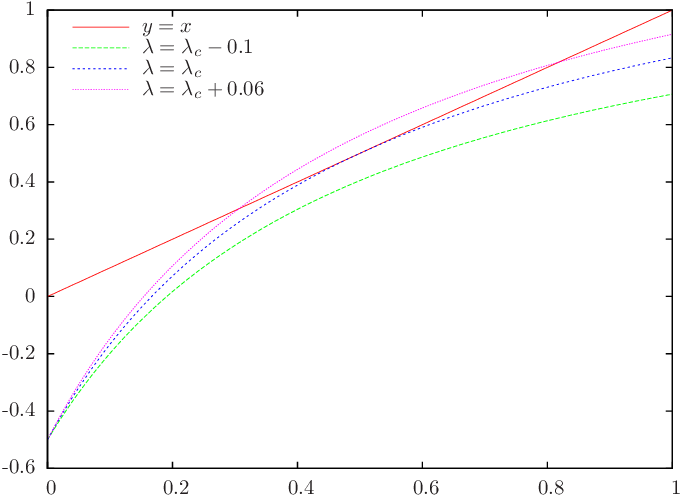} 
\hspace{1 cm} \includegraphics[width=7cm]{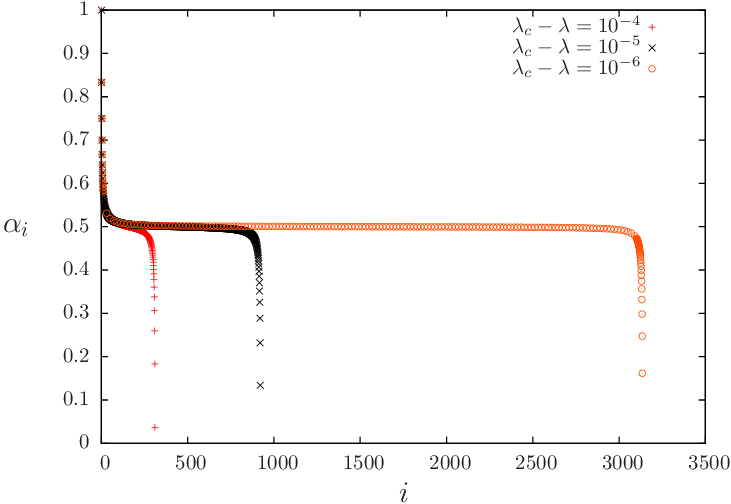}
\caption{Left panel: 
$f(x,\lambda)$ for different $\lambda$ above, at and below $\lambda_c$.
Right panel: 
the sequences $\alpha_i$ solution of Eq.~(\ref{eq_recursion_alpha}) for values
of $\lambda$ slightly smaller than $\lambda_c$. All data are for $k=2$.}
\label{fig_study_recurrence}
\end{figure}

For a given $p$ the supremum is over a smooth function of the $|\cE_p|$ reals
$x_e$; assuming that the supremum is reached in the interior of its domain, 
one has to look for the critical points
of the function. There is always one radially symmetric critical point,
i.e. where $x_e$ depends on $e$ only through the depth $i$ of the edge;
we denote $\hx_i$ the common value of $x_e$ for all edges at depth $i$.
We assume that this is the global maximum of the function.
This yields
\begin{equation}
\varkappa(n) \le 
\varkappa_*(n) =  \max_{p \in [1,n/2]} \sup_{\substack{\hx_1,\dots,\hx_p \ge 0 \\
2 \sum_{i=1}^p v_i \hx_i = n }} K_p(\hx_1,\dots,\hx_p) \ , 
\label{eq_def_varkappastar}
\end{equation}
where we defined the radially symmetric function $K_p$ as
\begin{equation}
K_p(\hx_1,\dots,\hx_p)=
\sum_{i=0}^{p-1} v_i [\ln \Gamma(\hx_i + k \hx_{i+1}) - \ln \Gamma(\hx_i) 
- k\ln\Gamma(1+\hx_{i+1})] \ .
\label{eq_def_Kp}
\end{equation}
For $i \ge 1$ we denoted $v_i=(k+1) k^{i-1}$ the number of edges of depth
$i$, and we set by convention $v_0=1$ and $\hx_0 = \hx_1 + 1$.
For a given $p$ the critical point of $K_p$
is achieved at the solution of:
\begin{equation} 
\psi(\hx_i+k \hx_{i+1}) - \psi(\hx_i) + \psi(\hx_{i-1}+k \hx_i)- \psi(\hx_i+1) 
= \lambda \ \quad \text{for} \ \ i \in [1,p] \ \text{with} \ \ 
\hx_0 =1+\hx_1 \ , \ \ \hx_{p+1}=0 \ ,
\label{eq_x_i}
\end{equation}
where $\lambda$ is a Lagrange multiplier that has to be fixed to the
value enforcing the constraint $\sum_{i=1}^p v_i \hx_i = n/2$, and 
$\psi \equiv \frac{ \dd \ln{\Gamma}}{\dd x}$ is the digamma function.
Note that $\varkappa_*$ is defined by extension for real values of $n$,
and that $\lambda$ plays the role of a parameter conjugated to $n$.
The previous equation can be rewritten as a three terms recurrence on $\hx_i$:
\begin{equation} 
\label{eq_recursion_x_i} 
\hx_{i+1} = \frac{1}{k} \left[ -\hx_i 
+ \psi^{-1} (\lambda + \psi(\hx_i) +\psi(\hx_i+1) - 
\psi(\hx_{i-1}+k \hx_i))\right] \ .
\end{equation}
We solved numerically the optimization problem defined in 
Eq.~(\ref{eq_def_varkappastar}); the plot in Fig.~\ref{fig_varkappa}
display the values of $\varkappa_*(n)/n$ we found in this way, as well as
the good agreement with our conjecture on its asymptotic behavior:
\begin{equation}
\frac{\varkappa_*(n)}{n}= \ln(2\sqrt{k})- \frac{(\pi \ln k)^2}{2(\ln n)^2} + 
o\left(\frac{1}{(\ln n)^2} \right) \ ,
\label{eq_conjecture_varkappastar}
\end{equation}
which implies the upperbound stated in Eq.~(\ref{eq_lemme_kappa}).
We turn now to the computation that led us to the formula 
in Eq.~(\ref{eq_conjecture_varkappastar}). In the large $n$ limit most
of the non-zero $\hx_i$ can be expected to be large (this is also confirmed
by the numerical computation of $\varkappa_*$). In consequence we can
simplify the recurrence relation (\ref{eq_recursion_x_i}) using the first
term of the asymptotic expansion of the digamma function, 
$\psi(x) = \ln(x) + O(1/x) $. The resulting relation is more compactly
written in terms of $\alpha_i = \frac{\hx_i}{\hx_{i-1}}$, which is found
to obey
\begin{equation} 
\alpha_{i+1} = f(\alpha_i, \lambda) \ , \quad \text{with} \ \ 
f(\alpha,\lambda) =
\frac{1}{k} \left(-1 + \frac{e^\lambda}{k+\frac{1}{\alpha}} \right) \ . 
\label{eq_recursion_alpha}
\end{equation} 
At the order of our approximation the initial condition 
$\alpha_1=\frac{\hx_1}{1+\hx_1}$ is equal to $\alpha_1=1$.
The fixed-point equation $\alpha=f(\alpha,\lambda)$ undergoes a bifurcation
transition at a critical value $\lc=\ln(4k)$ (see left panel
in Fig.~\ref{fig_study_recurrence}). For $\lambda>\lc$ there are
two solutions which coalesce in $\alpha_*=\frac{1}{k}$ at $\lc$,
and disappear for $\lambda<\lc$. This bifurcation has the following
consequences on the solution of Eq.~(\ref{eq_recursion_alpha}) with
initial condition $\alpha_1=1$, that for clarity we denote 
$\alpha_i(\lambda)$ to emphasize its dependency on the parameter $\lambda$:
\begin{itemize}
\item right at the transition the sequence decays towards its fixed-point
as $1/i$, more precisely
\begin{equation}
\alpha_i(\lc) = \frac{1}{k} + \frac{2}{k} \frac{1}{i} 
+ o\left( \frac{1}{i}\right) \ .
\label{eq_alpha_i_lc}
\end{equation}

\item when $\lambda \to \lc^-$ a plateau develops around the avoided 
fixed-point $\alpha_*=1/k$, with a diverging length of order 
$(\lc-\lambda)^{-1/2}$, see the right panel of 
Fig.~\ref{fig_study_recurrence}. One can establish the following scaling
behavior by zooming around the plateau:
\begin{equation}
\lim_{\lambda \to \lc^-}\left[ \frac{1}{\sqrt{\lc-\lambda}} 
\left( \alpha_{i=\frac{\theta}{\sqrt{\lc-\lambda}}}(\lambda) 
-\frac{1}{k}\right)
\right]
= -\frac{2}{k} \tan\left(\theta - \frac{\pi}{2} \right) \quad \forall \ \ 
\theta \in ]0,\pi[ \ .
\label{eq_alpha_i_scaling_precise}
\end{equation}
This in particular shows that the length of the plateau is 
$\pi/\sqrt{\lc - \lambda}$ at the leading order. We shall use this scaling 
behavior under the less precise but more readable form:
\begin{equation}
\alpha_i \approx \frac{1}{k} - \frac{2}{k} \sqrt{\lc - \lambda} 
\tan\left( i\sqrt{\lc -\lambda} - \frac{\pi}{2} \right) \ ,
\label{eq_alpha_i_scaling_approx}
\end{equation}
where it is understood that $i$ is in the scaling regime, i.e. of the
form $\theta/\sqrt{\lc-\lambda}$ with $\theta \in ]0,\pi[$.
\end{itemize}
Note that there is a matching condition between these two regimes: if one
considers a value of $i$ very large but finite with respect to 
$(\lc-\lambda)^{-1/2}$, the same behavior of $\alpha_i(\lc)$ is obtained 
by taking $i \to \infty$ in Eq.~(\ref{eq_alpha_i_lc}) or 
$\lambda \to \lc$ with $i$ fixed in Eq.~(\ref{eq_alpha_i_scaling_approx}).
The above behaviors are generic for all recursion of the forms 
$\alpha_{i+1}=f(\alpha_i,\lambda)$ that encounters a bifurcation transition,
and the quantitative statements only depend on the value of the partial 
derivatives $\partial_\lambda f$ and $\partial^2_{\alpha,\alpha} f$ of $f$
in $(\alpha_*,\lc)$. It turns out in addition that for the specific function
$f$ of Eq.~(\ref{eq_recursion_alpha}) the critical sequence can be computed
exactly, $\alpha_i(\lc)=((k+1)+(k-1)i)/(k(3-k)+k(k-1)i)$.

\begin{figure}
\center
\includegraphics[width=7cm]{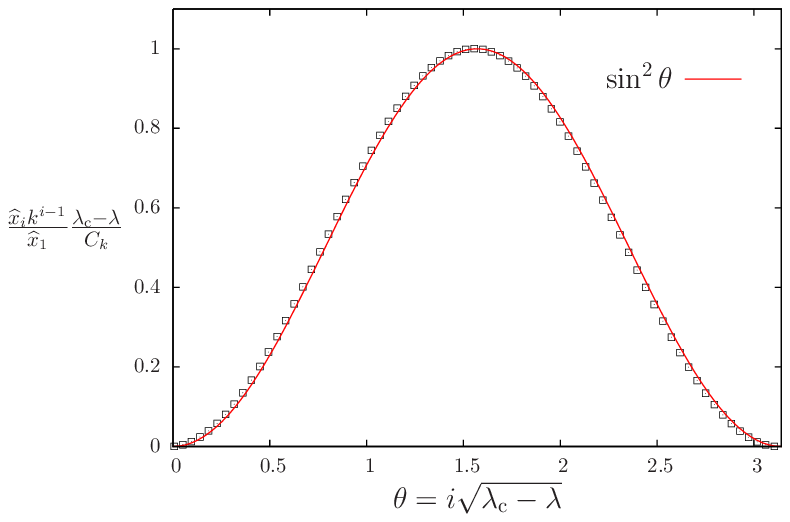}
\caption{Verification of Eq.~(\ref{eq_hxi_scaling}) for $k=2$, 
$\lambda = \lambda_{\rm c} - 4 . 10^{-5}$. The solid line is the analytical 
prediction, the symbols are the $\hx_i$ obtained from the resolution
of the optimization problem of Eq.~(\ref{eq_def_varkappastar}), and are 
indistinguishable on this scale from their approximate computation
in terms of the $\alpha_i$. Note that there is no fitting parameter in this 
figure.}
\label{fig_xi}
\end{figure}

Let us now investigate the consequence of these behaviors of 
$\alpha_i(\lambda)$ on the properties of the $\hx_i(\lambda)$.
The leading order of the recurrence relation (\ref{eq_recursion_x_i})
only constrains the ratio $\alpha_i$ between successive terms in the
sequence $\{\hx_i\}$, in consequence the initial condition $\hx_1$ is at 
this point a free parameter that we shall fix afterwards. By the definition 
of $\alpha_i$ we have
\begin{equation}
\hx_i(\lambda) = \hx_1 \left(\frac{1}{k}\right)^{i-1} 
\prod_{j=2}^i (k \alpha_j(\lambda)) \ ,
\end{equation}
where we have factorized the plateau value $1/k$ of the $\alpha_i$.
Consider first the large $i$ regime at the transition, i.e. for 
$\lambda = \lc$. As a consequence of (\ref{eq_alpha_i_lc}) one has
$\sum_{j=2}^i \ln(k\alpha_j(\lc)) = 2 \ln(i) + \ln(C_k) + O(1/i)$, hence
\begin{equation}
\hx_i(\lc) = \hx_1 \left(\frac{1}{k}\right)^{i-1} i^2 C_k (1+O(1/i)) \ ,
\label{eq_hxi_lc}
\end{equation}
where $C_k$ is a positive $k$-dependent constant. Actually one can compute 
$C_k$ explicitly thanks to the exact expression of $\alpha_i(\lc)$, and find
$C_k=(k-1)^2/(2k(k+1))$. Let us now turn to the
scaling regime $\lambda \to \lc^-$ where Eq.~(\ref{eq_alpha_i_scaling_approx})
is valid. Consider two indices $i_0 < i$ of the form 
$i_0=\theta_0/\sqrt{\lc-\lambda}$ and $i=\theta/\sqrt{\lc-\lambda}$. One has
\begin{eqnarray}
\hx_i(\lambda) &=& \hx_{i_0}(\lambda) \left(\frac{1}{k}\right)^{i-i_0}
\exp\left[\sum_{j=i_0+1}^i \ln(k\alpha_j(\lambda) ) \right] \\
&\approx&  \hx_{i_0}(\lambda) \left(\frac{1}{k}\right)^{i-i_0}
\exp\left[- 2 \sqrt{\lc - \lambda}\sum_{j=i_0+1}^i\tan\left( j\sqrt{\lc -\lambda} - \frac{\pi}{2} \right)  \right] \ ,
\end{eqnarray}
where we have used Eq.~(\ref{eq_alpha_i_scaling_approx}) in the second line.
Now in the limit $\lambda \to \lc$ the term in square brackets is the Riemann
discretization of the integral of $\tan(\theta - \pi/2)$, hence
\begin{equation}
\hx_i(\lambda) \approx \hx_{i_0}(\lambda) k^{i_0-1} \frac{1}{\sin^2(\theta_0)}
\left(\frac{1}{k}\right)^{i-1} \sin^2(\theta) \ .
\end{equation}
Finally we invoke a matching argument as explained above~: we take $i_0$
large but finite with respect to $(\lambda-\lc)^{-1/2}$, so that 
$\theta_0 \to 0$, and we replace the value of $\hx_{i_0}$ by its value computed
at $\lc$ in Eq.~(\ref{eq_hxi_lc}). This yields
\begin{equation}
\hx_i(\lambda) \approx \frac{\hx_1 C_k}{\lc-\lambda} 
\left(\frac{1}{k}\right)^{i-1} \sin^2(i \sqrt{\lc-\lambda}) \ ,
\label{eq_hxi_scaling}
\end{equation}
valid in the scaling regime $i=\theta/\sqrt{\lc-\lambda}$ with 
$\theta \in ]0,\pi[$. In Fig.~\ref{fig_xi} we present a numerical confirmation
of this prediction.

As can be seen on the right panel of Fig.~\ref{fig_study_recurrence},
once the the sequence $\alpha_i$ exits the scaling regime around the 
plateau a finite number of additional iterations brings it to negative 
values. Let us call $p(\lambda)$ the largest value of $i$ such that 
$\alpha_i(\lambda)$ is positive. From the observation above, 
$p(\lambda)=\pi/\sqrt{\lc-\lambda}$ at the leading order when 
$\lambda \to \lc^-$. Let us now discuss the scaling of $\hx_1$. For the
computation above to be consistent with the replacement of $\psi(x)$ by
$\ln(x)$ it was based upon, one should have $\hx_i(\lambda)$ large for most 
of the $i$'s between $1$ and $p(\lambda)$. On the other hand one should 
impose in an approximated way the boundary condition $\hx_{p(\lambda)+1}=0$
of the exact system (\ref{eq_x_i}). In consequence we shall choose
$\hx_1$ such that $\hx_i$ is of order one for $i$ at the end of the scaling 
regime ($i = \pi^-/\sqrt{\lc-\lambda}$), hence at the leading 
order (within multiplicative constant and sub-exponential corrections
in $(\lc-\lambda)^{-1/2}$)
\begin{equation}
\hx_1(\lambda) \approx k^{\frac{\pi}{\sqrt{\lc-\lambda}}} \ , \quad 
\text{i.e.} \qquad \lc - \lambda \sim \frac{(\pi \ln k)^2 }{(\ln \hx_1)^2} \ .
\label{eq_leading_hx1}
\end{equation}
This ends our determination of the asymptotic form of the optimal value
of the sequence $\{\hx_i\}_{i=1}^{p(\lambda)}$ as a function of $\lambda$;
from it we shall now compute the corresponding values of $n$ and 
$\varkappa_*$. The former reads
\begin{equation}
n(\lambda) =2 (k+1) \sum_{i=1}^{p(\lambda)} k^i \hx_i(\lambda) \approx
2 \frac{k+1}{k} \frac{C_k \hx_1(\lambda)}{(\lc-\lambda)^{3/2}} \int_0^\pi
\dd \theta \ \sin^2(\theta) =
\pi \frac{k+1}{k} \frac{C_k \hx_1(\lambda)}{(\lc-\lambda)^{3/2}} \ ,
\label{eq_n_approx}
\end{equation}
where we replaced the sum over $i$ by an integral over $\theta$ using the
form (\ref{eq_hxi_scaling}) of the $\hx_i$ in the scaling regime.
The leading behavior of $n$ is thus dictated by the one of $\hx_1(\lambda)$,
and the relation (\ref{eq_leading_hx1}) can thus be re-expressed as
\begin{equation}
\lc - \lambda \sim \frac{(\pi \ln k)^2 }{(\ln n)^2} \ .
\label{eq_lcml_n}
\end{equation}
To compute the value of $\varkappa_*$ achieved in the large $n$ (or
equivalently $\lambda \to \lc^-$) limit we use the asymptotic expansion
$\ln \Gamma(x)= x \ln x - x +O(\ln x)$ (consistently with our approximation
of $\psi(x)$), the term in square brackets in (\ref{eq_def_Kp}) reading
\begin{equation}
\hx_i [\ln(4k) + \beta_{i+1} \ln(2k) 
+ (2+\beta_{i+1}) \ln\left(1+ \frac{\beta_{i+1}}{2} \right)  
- (1+\beta_{i+1}) \ln(1+\beta_{i+1}) ] + O(\ln \hx_i) \ ,
\end{equation}
where we defined $\beta_i$ by $\alpha_i=\frac{1}{k}(1+\beta_i)$. Once inserted
in the sum over $i$ the first term leads to a factor proportional to $n$,
\begin{equation}
\varkappa_*(n) \approx n \ln(2 \sqrt{k}) + \sum_i v_i \hx_i [\beta_{i+1} \ln(2k) + (2+\beta_{i+1}) \ln\left(1+ \frac{\beta_{i+1}}{2} \right)  - (1+\beta_{i+1}) \ln(1+\beta_{i+1}) ] \ .
\end{equation}
In the scaling regime 
$\beta_i=-2\sqrt{\lc-\lambda} \tan(i\sqrt{\lc - \lambda} - \frac{\pi}{2})$ 
is small, we thus expand the previous expression to second order in $\beta_i$
and trade the sum with an integral to obtain 
\begin{equation}
\varkappa_*(n) \approx n \ln(2 \sqrt{k})  + \frac{k+1}{k} 
\frac{\hx_1(\lambda) C_k}{(\lc-\lambda)^{3/2}} 
\int_0^\pi \dd \theta \, \sin^2(\theta) 
\left[-2\ln(2 k) \sqrt{\lc-\lambda} \tan\left(\theta - \frac{\pi}{2}\right) - 
(\lc-\lambda)^2 \tan^2\left(\theta - \frac{\pi}{2}\right) \right] \ .
\end{equation}
The integral of the first term vanishes for symmetry reasons, the second
one is easily evaluated, and by comparison with the expression of $n$
given in (\ref{eq_n_approx}) one obtains
\begin{equation}
\frac{\varkappa_*(n)}{n} \approx 
\ln(2 \sqrt{k}) - \frac{1}{2} (\lc - \lambda) \ .
\end{equation}
With the expression of $\lc-\lambda$ as a function of $n$ given in 
(\ref{eq_lcml_n}), this completes the derivation of 
(\ref{eq_conjecture_varkappastar}).

Finally the lowerbound in (\ref{eq_lemme_kappa}) can easily be justified by
contradiction: suppose there exists $x>0$ such that for arbitrarily large
$n$, $\varkappa(n) \leq (2 \sqrt{k})^n e^{- \frac{n}{(\ln n)^{2-x}}}$.
Then one could repeat the derivation of Sec.~\ref{sec_bounds_offd_upper},
choosing $r_{\rm m}(n)=\lfloor \frac{n}{(\ln n)^{2-\frac{x}{2}}}\rfloor$, and
obtain instead of (\ref{eq_upper_bound_off_diag_proof}) the upperbound
\begin{equation}
\frac{1}{n} \ln (c_n) \le \ln(2 \sqrt{k}) 
- \frac{1}{(\ln n)^{2-\frac{x}{2}}} \ln (\E[J^2]^{-1}) \ .
\end{equation}
This would violate the lowerbound in (\ref{eq_statement_offd}), hence the
contradiction.

\section{Proofs for the diagonal disorder model}
\label{sec_bounds_d}
This section contains the proofs of the bounds stated in 
Eq.~(\ref{eq_statement_d}) for the moments $d_n$ of the diagonal disorder 
case, i.e. for edge couplings $J_{ij}$ deterministically
equal to $1$ and vertex random variables $V_i$ drawn in $[0,W]$ with a
probability distribution satisfying the assumptions (D1-D2) of 
Sec.~\ref{sec_def}. The proofs are similar to the off-diagonal case,
we mostly precise the additional technicalities that arises here.

\subsection{Lower bound}
\label{sec_bounds_d_lower}

As in the off-diagonal case the lower bound on $d_n$ follows by considering
only the walks restricted to a neighborhood of depth $h$ around the root of
the tree, and by conditioning on the event that the disorder is large in
this neighborhood.

We recall that $d_n = \sum_{\w \in \W_n} \pi(\w)$, where $\W_n$ denotes now
the set of closed walks on $\T_k$ of length $n$ with self-bond steps allowed, 
and the weight of a walk is 
$\pi(\omega) = \E[\prod_{v \in \cV} V_v^{s_v(\omega)} ]$,
with $s_v(\omega)$ the number of self-bond steps taken by the walk around
the vertex $v$ of the tree. These weights being positive we can lowerbound
$d_n$ as
\begin{equation}
d_n \ge \sum_{\w \in \tW_{n,h}} \pi(\omega) \ ,
\end{equation}
where $\tW_{n,h}$ is the set of walks on $\tT_k$ that visit vertices at
distance at most $h$ from the root. We call $\cV_h$ the set of such vertices,
introduce a positive threshold $V_0<W$, and define the event $M$ as
$\{V_v \ge V_0 \ \forall v \in \cV_h\}$. With a reasoning similar
to the one which yielded Eq.~(\ref{eq_offd_bound_pi}) we obtain
\begin{equation}
\pi(\w) \ge \P[V \ge V_0]^{|\cV_h|} \, V_0^{\stot(\w)}
\end{equation}
for all walks $\w \in \tW_{n,h}$, where we defined $\stot(\w) = \sum_v s_v(\w)$
the total number of self-bond steps taken by the walk. Putting these two
inequalities together one has
\begin{equation}
d_n \ge \P[V \ge V_0]^{|\cV_h|} \sum_{\w \in \tW_{n,h}} V_0^{\stot(\w)} \ .
\end{equation}
We now use the projection from a walk to its depth first introduced in
Sec.~\ref{sec_pure_case}. The self-bond steps are then associated to 
horizontal steps in the path, which is called a Motzkin path in this case.
The sum over $\w$ in the last equation is thus equal to the weighted 
sum over Motzkin path of length $n$, where ascending steps have weight $k$
(the number of branches the walk can choose from in a step away from the root) 
and horizontal (resp. descending)
steps have weight $V_0$ (resp. $1$). This quantity $m_{n,h}$ is computed
by combinatorial techniques in App.~\ref{app_lower_bound}. The number
of vertices at depth at most $h$ is $|\cV_h|=\frac{k^{h+1}}{k-1}$, hence
\begin{equation}
\frac{1}{n} \ln d_n \ge \frac{1}{n} \ln m_{n,h}
+ \frac{k^{h+1}}{n (k-1)}\ln \P[V \ge V_0] \ .
\label{eq_d_lb}
\end{equation}
If we let $n\to \infty$ with $h$ and $V_0$ fixed, the asymptotic properties
of $m_{n,h}$ proved in App.~\ref{app_lower_bound} 
(see Eq.~(\ref{eq_mnh_fixed})) yields 
\begin{equation}
\liminf_{n \to \infty} \frac{1}{n} \ln d_n  \ge \ln
\left(V_0 + 2\sqrt{k} \cos\left(\frac{\pi}{h+2} \right) \right) \ .
\end{equation}
Letting finally $h \to \infty$ and $V_0 \to W$ and noting that $d_n \le d_n^0$,
the shifted pure model with $J_{ij}=1$ and $V_i=W$ for which 
$\lim \frac{1}{n}\ln d_n^0 = \ln (2 \sqrt{k} + W )$ allows to conclude
$\lim \frac{1}{n}\ln d_n = \ln (2 \sqrt{k} + W )$. The upper limit of the
support is thus $2 \sqrt{k} + W$ in the diagonal disorder case, in agreement
with the ergodicity arguments sketched in Sec.~\ref{sec_def_model_infinite}.

We now consider the limit $n\to \infty$, taking the parameters 
$h \to \infty$ and $V_0 \to W$ in an $n$-dependent way. For the
same reasons as in the off-diagonal case the optimal scaling of the
parameters is $h=\lfloor\alpha \ln n \rfloor$ 
with $\alpha< 1/\ln k$ and 
$V_0=W-\frac{1}{(\ln n)^{2+y}}$ where $y>0$ is an arbitrary positive constant.
Using the asymptotic expansion (\ref{eq_mnh_variable}) for $m_{n,h}$, 
the r.h.s. of (\ref{eq_d_lb}) becomes
\begin{equation}
\ln(2\sqrt{k}+W) - 
\frac{\pi^2 \sqrt{k}}{\alpha^2(2\sqrt{k}+W)} \frac{1}{(\ln n)^2}
+\frac{k}{k-1}\frac{1}{n^{1-\alpha \ln k}} 
\ln \P\left[V \ge  W-\frac{1}{(\ln n)^{2+y}}\right]
+O\left(\frac{1}{(\ln n)^{2+y}} \right) \ .
\end{equation}
The assumption (D2) on the random variable $V$ ensures that the third term
is negligible for any $\alpha<1/\ln k$ and $y>0$,
which concludes the proof of the lower bound in Eq.~(\ref{eq_statement_d}).

\subsection{Upper bound}
\label{sec_bounds_d_upper}

We present now the proof of the upper bound in  Eq.~(\ref{eq_statement_d}).
Let us first recall the definition of the moment of the shifted pure case 
($J_{ij}=1$, $V_i=W$) in terms of the weights $\pi^0(\w)$ of the walks:
\begin{equation}
d_n^0 = \sum_{\w \in \W_n} \pi^0(\w) \ , \qquad
\pi^0(\w) = \prod_{v \in \cV} W^{s_v(\w)} = W^{s_{\rm tot}(\w)} \ ,
\end{equation}
$d_n^0$ being equal to $(2\sqrt{k}+W)^n$ within polynomial corrections.
Consider now the effect of the disorder on the weight of a walk:
\begin{equation}
\pi(\w) = \prod_{v \in \cV} \E[V^{s_v(\w)}] = \pi^0(\w) 
\prod_{v \in \cV} \E\left[\left(\frac{V}{W} \right)^{s_v(\w)} \right]
\le \pi^0(\w) \left(\E\left[\frac{V}{W} \right]\right)^{\tau(\w)} \ ,
\end{equation}
where $\tau(\w) = |\{v \in \cV | s_v(\w) \ge 1\} $  is the size of the 
self-support of the walk, as defined in Sec.~\ref{sec_walks_on_the_tree}.
By our assumption (D1) $\E[V/W]$ is strictly smaller than 1, hence the
weight of a walk is exponentially suppressed in the size of its self-support.
The proof will thus be based on partitioning the set of walks in $\W_n$ between
those with a large self-support, for which we shall use the bound above, and
on bounding the contribution of the walks with small self-supports.

Let us introduce a growing integer function $t_{\rm m}(n)$, with 
$t_{\rm m}(n) \to \infty$ and $\frac{t_{\rm m}(n)}{n} \to 0$ as $n \to \infty$,
and denote $\W_n^\le$ the subset of $\W_n$ which contains the walks whose 
self-support contains less than $t_{\rm m}(n)$ vertices,
$\W_n^\le = \{\w \in \W_n | \tau(\w) \le t_{\rm m}(n) \}$. Then
\begin{equation}
d_n \le \left(\E\left[\frac{V}{W} \right]\right)^{t_{\rm m}(n)} 
\sum_{\w \in \W_n  \setminus \W_n^\le} \pi^0(\w) + \sum_{\w \in \W_n^\le} \pi^0(\w)
\le  \left(\E\left[\frac{V}{W} \right]\right)^{t_{\rm m}(n)} d_n^0 +
\sum_{\w \in \W_n^\le} \pi^0(\w) \ ,
\end{equation}
where we have used the obvious facts that $\pi(\w) \le \pi^0(\w)$ and
$\pi^0(\w) \ge 0$ for all $\w$.

We shall further partition $\W_n^\le$ according to the size of the supports
of the walks (i.e. the number of edges which are crossed at least once by
the walk). Let us introduce another growing threshold function $r_{\rm m}(n)$
with $r_{\rm m}(n) \to \infty$ and $\frac{r_{\rm m}(n)}{n} \to 0 $, and
define $\W_n^1 = \{ \w \in \W_n^\le | |\sigma(\w)| \le r_{\rm m}(n)  \}$ and
$\W_n^2 = \{ \w \in \W_n^\le | |\sigma(\w)| > r_{\rm m}(n)  \}$.
We can thus refine the bound above as
\begin{equation}
d_n \le \left(\E\left[\frac{V}{W} \right]\right)^{t_{\rm m}(n)} d_n^0 +
\sum_{\w \in \W_n^1} \pi^0(\w) + \sum_{\w \in \W_n^2} \pi^0(\w) \ .
\label{eq_upper_bound_separation}
\end{equation}

We shall work out separately upperbounds for the contributions of $\W_n^1$ and
$\W_n^2$. Let us start with the former, and introduce a larger set of walks
$\W_n^{1'} = \{ \w \in \W_n | |\sigma(\w)| \le r_{\rm m}(n)  \}$, in which
we have removed the constraint on the size of the self-supports. By the
positivity of the weights the sum over $\W_n^{1'}$ dominates the one on 
$\W_n^1$.
The crucial point is now to realize that there is a ${n \choose s}$ to 1 mapping
between each walk $\w \in \W_n$ with $s_{\rm tot}(\w)=s$ and the walk 
$\ow \in \M_{n-s}$ obtained by removing all self-bond steps, that in consequence
shares the same skeleton. This binomial coefficient counts the number of
possible choices for
the $s$ times $0 \le t_1 \le \dots \le t_s \le n-s$ at which a self-bond step 
is inserted in $\ow$ to construct $\w$. For completeness we give in 
App.~\ref{app_nstoone} a more formal proof of this fact based on the study of 
the combinatorial factors. We can thus write
\begin{equation}
\sum_{\w \in \W_n^1} \pi^0(\w) \le \sum_{\w \in \W_n^{1'}}  \pi^0(\w) = \sum_{s=0}^n
{n \choose s} W^s |\M_{n-s}^\le| \ ,
\end{equation}
where $\M_{n-s}^\le$ is the set of walks of length $n-s$, without self-bond 
steps, studied in Sec.~\ref{sec_bounds_offd_upper}. We shall thus exploit
the bound on $|\M_n^\le|$ derived in that section: we choose an arbitrary $x>0$,
$\beta < (\ln k)^2\pi^2/2$ and set 
$r_{\rm m}(n)= \lfloor \frac{n}{(\ln n)^{2+x}} \rfloor$. From 
Eq.~(\ref{eq_domination_grands_squelettes}) we obtain the existence of an 
(even) $n_0$
such that, for $n-s \ge n_0$,
\begin{equation}
|\M_{n-s}^\le| \leq (2 \sqrt{k})^{n-s} e^{-\beta \frac{n-s}{(\ln(n-s))^2}} \le
(2 \sqrt{k})^{n-s} e^{-\beta \frac{n-s}{(\ln n)^{2}}} \ ,
\end{equation}
this bound being trivially true for odd values of $n-s$ for which
$|\M_{n-s}^\le|=0$. Inserting this bound in the inequality above yields
\begin{eqnarray}
\sum_{\w \in \W_n^1} \pi^0(\w) &\le& \sum_{s=0}^{n-n_0}{n \choose s} W^s 
(2 \sqrt{k})^{n-s} e^{-\beta \frac{n-s}{(\ln n)^{2}}}
+ \sum_{s=n-n_0+1}^n{n \choose s} W^s |\M_{n-s}^\le| \\ 
&\le& \sum_{s=0}^n{n \choose s} W^s 
(2 \sqrt{k})^{n-s} e^{-\beta \frac{n-s}{(\ln n)^{2}}}
+ |\M^\le_{n_0}| \sum_{s=0}^n{n \choose s} W^s \\
&=& e^{-\beta \frac{n}{(\ln n)^{2}}} 
(2 \sqrt{k} + W e^{\beta \frac{1}{(\ln n)^{2}}})^n + |\M^\le_{n_0}| (1+W)^n \ .
\end{eqnarray}
From the first to the second line we have used the fact that $|\M^\le_n|$ grows 
with (even) $n$ and extended the range of the sums, all terms being 
non-negative.
As $n_0$ is now fixed and $2\sqrt{k} > 1$ the second term is exponentially
smaller than the first one in the large $n$ limit. Moreover the first one
can be expanded as
\begin{eqnarray}
e^{-\beta \frac{n}{(\ln n)^{2}}} 
(2 \sqrt{k} + W e^{\beta \frac{1}{(\ln n)^{2}}})^n &=& e^{-\beta \frac{n}{(\ln n)^{2}}}
(2 \sqrt{k} + W)^n \left(1 + \frac{W}{2\sqrt{k}+W} \left(e^{\beta \frac{1}{(\ln n)^{2}}} -1 \right) \right)^n \\
&=& (2 \sqrt{k} + W)^n 
e^{- \beta \frac{2\sqrt{k}}{2 \sqrt{k}+W} \frac{n}{(\ln n)^2} + O(n/(\ln n)^4)} \ .
\end{eqnarray}
We can thus conclude the study of the sum over $\W_n^1$: for all
$\gamma<\frac{(\pi \ln k)^2}{2} \frac{2\sqrt{k}}{2\sqrt{k}+W}$ it holds
for $n$ large enough that
\begin{equation}
\frac{1}{d_n^0} \sum_{\w \in W_n^1} \pi^0(\w) \le e^{-\gamma \frac{n}{(\ln n)^2}} 
\ .
\label{eq_upper_bound_W1}
\end{equation}

Let us finally derive an upperbound on the contributions of the walks in the
set $\W_n^2$ whose support contains more than $r_{\rm m}(n)$ edges while their
self-support contains less than $t_{\rm m}(n)$ vertices. We write this sum as
\begin{equation}
\sum_{\w \in \W_n^2} \pi^0(\w) = 
\sum_{s=0}^{n-2r_{\rm m}(n)-2} W^s \sum_{r=r_{\rm m}(n)+1}^{n/2}
\ \sum_{\ow \in \M_{n-s}^r} \
\sum_{\substack{\cA \subset \sigma(\ow)\cup 0 \\ |\cA| \le t_{\rm m}(n)}} 
E(\ow,s,\cA) \ ,
\label{eq_sum_W2}
\end{equation}
where $\M_n^r$ is the set of walks of length $n$
without self-bond steps with a support of $r$ edges, and $E(\ow,s,\cA)$ is the
number of walks $\w \in \W_n$ that are obtained from $\ow \in \M_{n-s}$ by 
adding to it $s$ self-bond steps, in such a way that the self-support of $\w$ 
is a given $\cA$, a subset of the vertices visited by $\ow$. 
A short combinatorial reasoning presented in App.~\ref{app_nstoone} yields
the following upperbound for this quantity
\begin{equation}
E(\ow,s,\cA) \le {n-2|\sigma(\ow)| + (k+1) |\cA| \choose s}  \ .
\label{eq_ub_E}
\end{equation}
In all the terms of (\ref{eq_sum_W2}) we have $|\sigma(\ow)|\ge r_{\rm m}(n)+1$
and $|\cA| \le t_{\rm m}(n)$; moreover the number of terms in the sum over
$\cA$ can be coarsely upperbounded by $t_{\rm m}(n) {n \choose t_{\rm m}(n)}$.
This being done one can relax the constraint over $r$ and sum over all walks
$\ow \in \M_{n-s}$, which are no more numerous than $(2\sqrt{k})^{n-s}$. This
yields
\begin{equation}
\sum_{\w \in \W_n^2} \pi^0(\w) \le
\sum_{s=0}^{n-2r_{\rm m}(n)-2} W^s (2\sqrt{k})^{n-s} 
t_{\rm m}(n) {n \choose t_{\rm m}(n)} 
{n-2 r_{\rm m}(n)-2 + (k+1) t_{\rm m}(n)  \choose s} \ .
\end{equation}
We choose now the threshold function on the self-support size as
$t_{\rm m}(n) = \lfloor \frac{n}{(\ln n)^{2+2x}}\rfloor$, which is in consequence
negligible with respect to $r_{\rm m}(n)$ as $n \to \infty$.
In order to perform the sum over $s$ we decompose the last binomial coefficient
as
\begin{eqnarray}
{n-2 r_{\rm m}(n)-2 + (k+1) t_{\rm m}(n)  \choose s} &=& 
{n-2 r_{\rm m}(n)-2  \choose s} \prod_{i=1}^{(k+1) t_{\rm m}(n) } 
\frac{n-2 r_{\rm m}(n)-2 + i}{n-2 r_{\rm m}(n)-2 -s + i} \\
&\le&  {n-2 r_{\rm m}(n)-2  \choose s} 
\frac{n^{(k+1)t_{\rm m}(n) }}{((k+1)t_{\rm m}(n))!} \ .
\end{eqnarray}
This yields
\begin{equation}
\sum_{\w \in \W_n^2} \pi^0(\w) \le t_{\rm m}(n) {n \choose t_{\rm m}(n)}
\frac{n^{(k+1)t_{\rm m}(n) }}{((k+1)t_{\rm m}(n))!} (2\sqrt{k} + W)^n 
\left(\frac{2\sqrt{k}}{2\sqrt{k} +W} \right)^{2 r_{\rm m}(n)+2} \ .
\end{equation}
Using standard bounds on factorials and binomial coefficients and
the fact that $t_{\rm m} \ll r_{\rm m}$, 
we can conclude on the existence of $\mu > 0$ such that for $n$ large
enough
\begin{equation}
\label{eq_upper_bound_W2} 
\frac{1}{d^0_n} \sum_{\w \in \W_n^2} \pi^0(\omega) \leq  
e^{- \mu \frac{n}{(\ln n)^{2+x}}} \ . 
\end{equation}

Finally, putting together (\ref{eq_upper_bound_separation}), (\ref{eq_upper_bound_W1}) and (\ref{eq_upper_bound_W2}), we obtain for $n$ large enough:
\begin{equation} \frac{d_n}{d^0_n} \leq 
\left(\E\left[\frac{V}{W} \right]\right)^{
\lfloor \frac{n}{(\ln n)^{2+2x}}\rfloor}
+ e^{- \gamma \frac{n}{(\ln n)^2}} +e^{- \mu \frac{n}{(\ln n)^{2+x}}} \ .
\end{equation}
The first term dominates when $n \to \infty$, hence we conclude that (renaming
$2x$ in $x$) for all $x >0$ and $n$ large enough :
\begin{equation} \frac{1}{n}\ln(d_n) \leq \ln(2 \sqrt{k}+W) 
- \frac{1}{(\ln n)^{2+x}} \ln(\E[V/W]^{-1}) \ .
\end{equation}
The same remarks as those following (\ref{eq_upper_bound_off_diag_proof}) 
hold, namely that the positive constant $\ln(\E[V/W]^{-1})$ is irrelevant
because of the condition $x>0$, and that an improvement of the proof 
would be in any case limited by the requirement
$\gamma < \frac{(\pi \ln k)^2}{2}\frac{2\sqrt{k}}{2 \sqrt{k}+W}$, in agreement
with the lower bound in Eq.~(\ref{eq_statement_d}).

\section{Conclusions}
\label{sec_conclusions}

\begin{figure}
\center
\includegraphics[height=7 cm]{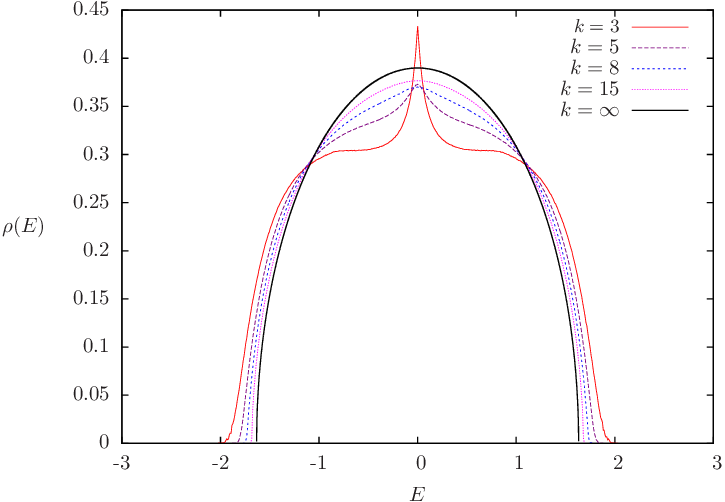} \hspace{1 cm}
\caption{The density of states in presence of off-diagonal disorder for 
different degrees, with rescaled couplings $J_{ij}$ uniformly random on
$[-\sqrt{2/k},\sqrt{2/k}]$.
The $k = \infty$ curve corresponds to the Wigner semi-circle law
of support $[-\sqrt{8/3},\sqrt{8/3}]$.}
\label{fig_differents_k}
\end{figure}

In this paper we have characterized the double-exponential Lifshitz behavior
of the Bethe lattice density of states close to its edge in presence of 
bounded (diagonal or off-diagonal) disorder in an indirect way, by 
controlling the asymptotic growth of its moments. As a byproduct of this 
study we have unveiled some
geometric properties of the dominant supports of closed random walks on
regular trees. Let us conclude by making a series of observations and 
suggestions for future work.

We believe that most of the methods and results of the paper could be
adapted to characterize the integrated density of states of the adjacency 
matrix of random graphs with arbitrary bounded degree distribution. Denoting
$k_{\rm max}$ the maximum degree, one expects a Lifshitz tail phenomenon to
occur around $2\sqrt{k_{\rm max}}$ as soon as the random graph is not
$k_{\rm max}$-regular. To characterize it one could use the results of the
off-diagonal disorder model on $\T_{k_{\rm max}-1}$, with $J_{ij}\in\{0,1\}$.
The $J_{ij}$ are i.i.d. only in the case of percolation (i.e. for a binomial
degree distribution), but in the general case two $J_{ij}$ are correlated 
only when they share a common vertex, so that it should be possible to handle
these weak correlations, thus extending the results 
of~\cite{muller,reinhold}.

When looking at the plot of the off-diagonal disorder density of states in
Fig.~\ref{fig_numerical_simulation} one could be tempted to divide its support
in two regions, one where it is ``large'' and one where the Lifshitz phenomenon
strongly reduces it. However the density of states is strictly positive 
everywhere inside its support, so that there is no unambiguous criterion
on the location of the frontier between these two regimes. Such a distinction
can be sharply defined by letting another parameter of the model evolve,
i.e. the degree $k+1$ of the lattice. In Fig.~\ref{fig_differents_k}
we have plotted the density of states of the off-diagonal disordered model
for various values of $k$; to allow for a meaningful comparison of these
curves we have scaled in accordance the magnitude of the disorder, taken
to be uniform on $[-\sqrt{2/k},\sqrt{2/k}]$. The support of the density
of states is thus, independently of $k$, the interval $[-2\sqrt{2},2\sqrt{2}]$.
However the density of states converges (pointwise) in this limit to the Wigner
semi-circle law whose support is controlled by the variance of the $J_{ij}$
(and not by their maximal value as the support of the density of states),
and which is found here to be $[-\sqrt{8/3},\sqrt{8/3}]$.
The regime $|E| \in[\sqrt{8/3} ,2\sqrt{2}]$ is
thus, in this $k \to \infty$ limit, the one of the Lifshitz tail.
Note the similarity with the unbounded (Gaussian or Cauchy) diagonal
disorder case studied in~\cite{AcKl92,MiDe94,AW10,AW11}, with $1/k$ playing
here the same role as the multiplicative constant in front of the unbounded 
$V_i$'s in these studies.

A possible direction for future studies would be the investigation of the
distribution of the largest eigenvalue of the Anderson model defined on 
random regular graphs, more precisely of its deviation from the upper limit
of the density of states. In the Gaussian ensembles of random matrix theory
the density of states is the Wigner semi-circle law, which vanishes as a 
square root at its edge; in that case the typical fluctuations of the largest 
eigenvalue around the edge of the density of states
are of order $N^{-2/3}$ and described by the Tracy-Widom law~\cite{TW96}. 
This result has been extended
by Sodin to the case of random regular graphs of fixed degree with 
$J_{ij}=\pm 1$~\cite{Sodin_journal,Sodin_preprint}. In presence of a Lifhsitz
tail in the density of states both the scaling with $N$
and the distribution of the fluctuations of the largest 
eigenvalue should be modified.

Finally, another direction of investigation could concern the numerical 
procedures used to solve Recursive Distributional Equations as (\ref{rde}).
The sample representation at the basis of the population 
dynamics~\cite{ACTA73,MePa_Bethe} is very natural and allows simple and
versatile implementations of the method, yet it performs poorly in sampling
rare events of the Lifshitz tail type. A combination between a sample 
representation of the typical part of the distribution and a mesh 
representation for its very small probability part might provide a better
alternative in such cases.

\acknowledgments

This work has been partly funded by the ANR project BOOLE, ANR-09-BLAN6011.
We warmly thank Giulio Biroli, Charles Bordenave, Marc Lelarge, Justin Salez, 
Sasha Sodin and Marco Tarzia for very useful discussions.

\appendix

\section{Numerical determination of the density of states and the 
center of the band singularity}
\label{appendix_middle}

Let us first explain the numerical procedure we followed to obtain the
density of states displayed in 
Figs.~\ref{fig_numerical_simulation},\ref{fig_cusp} and \ref{fig_differents_k}.
As shown in Sec.~\ref{sec_def_model_infinite} this computation amounts
to solve the Recursive Distributional Equation (RDE) of Eq.~(\ref{rde}), and
then to compute the average given in (\ref{densite_resolvent}).
The numerical procedure we implemented to solve this problem, known
as population dynamics~\cite{ACTA73,MePa_Bethe} 
or pool method~\cite{MoGa09}, consists
in approximating the distribution of $\widetilde{G}$ by the empirical 
distribution over a sample of ${\cal N} \gg 1$ (in practice we used 
${\cal N}$ of the order of $10^4$)
representative values of $\widetilde{G}$. This sample is updated according
to Eq.~(\ref{rde}) until numerical convergence is reached, then the average
in (\ref{densite_resolvent}) is estimated as an empirical average over the
sample (and over several update steps, a few hundreds for the data shown,
to increase the precision).
The numerical accuracy of this method is controlled by the size $\cal N$ of the
sample. The limit on the accessible values of $\cal N$ put by the memory
available on present computers is such that the regime of Lifshitz tails,
where the density of states is extremely small and hence requires a huge
precision to be determined, is not reachable via such a numerical procedure.

We shall now present the heuristic arguments which led us to the conjecture
that in the case of continuous off-diagonal disorder with support intersecting a neighbourhood of $0$, the density of states exhibit
a singularity at $E=0$ of the form
\begin{equation}
\rho(E) \sim \rho(0) - \alpha |E|^\frac{k-1}{2} \quad \text{as} \ \ E \to 0 \ ,
\end{equation}
where $\alpha$ is a positive constant. This conjecture is in very good
agreement with our numerical results for $k=2,3,4$, as shown explicitly
for $k=2$ in the right panel of Fig.~\ref{fig_cusp}.

Consider the RDE~(\ref{rde}) on $\widetilde{G}$ at the center of the band 
($E=0$), and in the limit of vanishing regularizing imaginary part $\eta$
(a similar limit of the RDE has been studied rigorously in~\cite{rank},
with $J_i=1$ and $k$ a Poisson random variable).
To simplify notations we introduce the random variable 
$\tg \equiv -i \widetilde{G}(0)$, solution of the RDE
\begin{equation}
\tg \overset{\textnormal d}{=} \frac{1}{\sum_{i=1}^k J_i^2 \tg_i} \ .
\label{rde_g}
\end{equation}
Since for $\textnormal{Im}\, z>0$ the relevant solution of the RDE~(\ref{rde})
has $\textnormal{Im}\, \widetilde{G}(z)>0$, we shall look for a solution
of (\ref{rde_g}) supported on $[0,\infty[$. 
We will assume that $\tg$ has a density $\widetilde{\mu}$ that decays as a 
power law for large values of $\tg$,
\begin{equation}
\label{levy} 
\widetilde{\mu}(\tg) \underset{\tg\rightarrow \infty}{\simeq} \tg^{-x-1} \ ,
\end{equation}
where in this appendix $\simeq$ means asymptotic equivalence within a 
multiplicative constant, and $x$ is an exponent that we shall determine
self-consistently. First note that (\ref{levy}) is equivalent to 
\begin{equation}
\label{levy2} 
\P(\tg \geq A) \underset{A \rightarrow \infty}{\simeq} \frac{1}{A^x} \ .
\end{equation}
The recursive distributional equation (\ref{rde}) relates the assumption on
the behavior of $\widetilde{\mu}$ near $\infty$ to its behavior near $0$: 
\begin{equation}
\begin{split} 
\P\left(\tg \leq a\right) &= 
\P \left (\sum_{i=1}^k J_i^2 \tg_i \geq \frac{1}{a} \right) 
\\ & \underset{a \rightarrow 0}{\simeq}\P \left(J^2 \tg \geq \frac{1}{a} \right)
\\ & \underset{a \rightarrow 0}{\simeq}\P \left(\tg \geq \frac{1}{a} \right) 
\underset{a\rightarrow 0}{\simeq} a^x \ .
\end{split} 
\end{equation}
In the first step we have used the heavy tail character of $\tg$, hence the
probability that the sum is large is close to the probability that a single
term is large, the second step relies on the boundedness of $J$, and
the last one follows from (\ref{levy2}).
Now we use again (\ref{rde}) in the other direction, i.e. to deduce the behavior
at large $\tg$ from the one at small $\tg$:
\begin{equation}
\begin{split} 
\P(\tg \geq A) &= \P\left( \sum_{i=1}^k J_i^2 \tg_i \leq \frac{1}{A} \right) 
\\ &\underset{A \rightarrow \infty}{\simeq} 
\left[ \P\left(J^2 \tg \leq \frac{1}{A} \right) \right]^k
\\ &\underset{A \rightarrow \infty}{\simeq}  \left[ 
\P\left(J^2 \leq \frac{1}{A}\right) 
\right]^k
\\ &\underset{A \rightarrow \infty}{\simeq} 
\left[ \frac{1}{\sqrt{A}}
\right]^k \ .
\end{split}\end{equation}
Indeed the probability that the sum in the first line is small is roughly
the probability that all the (positive) terms are small. In the second
step we assumed $x>1/2$, hence the probability that the product $J^2 \tg$
is small is controlled by the smallness of $J^2$, and the last step
follows from the assumption that $J$ has a finite density 
near $0$ (note that if the support of $J$ were bounded away from 0 there would
not be any singularity in the density of states). 
Comparing this with (\ref{levy2}) gives the self-consistency 
condition $x=k/2$, which confirms the hypothesis $x>1/2$ for all relevant 
$k$. The same computations applied to the density $\mu$ of $g\equiv -i G(0)$ 
yield the expansion: 
\begin{equation} 
\mu(g) \underset{g \rightarrow \infty}{\simeq} g^{-\frac{k+1}{2}-1} 
\sim \alpha' \, g^{-\frac{k+1}{2}-1} \hspace{1 cm} (\alpha' \in \mathbb{R}_+)
\ .
\end{equation}
Now we reintroduce the regularizing imaginary part $\eta$, and observe
that when it is small its main effect is to provide a cutoff at $1/\eta$
on the distribution of $\tg$ and $g$ determined above directly at $\eta=0$.
This yields
\begin{equation}
\begin{split} 
\label{asym_im} 
\frac{1}{\pi} \textnormal{Im} \mathbb{E} [G(0+i\eta)] 
&\underset{\eta \to 0}{\simeq} \frac{1}{\pi}\int_0^{1/\eta} g \mu(g) \dd g \\
&= \frac{1}{\pi}\int_0^\infty g \mu(g) \dd g 
- \frac{\alpha'}{\pi} \int_{1/\eta}^\infty g^{-\frac{k+1}{2}}
\left(1+o(1)\right) \dd g
=\rho(0)- \alpha \, \eta^{\frac{k-1}{2}} + o\left(\eta^{\frac{k-1}{2}}\right) \ .
\end{split} 
\end{equation}
The finiteness of $\rho(0)$ follows from the integrability of $g \mu(g)$,
and $\alpha$ is a positive constant proportional to $\alpha'$.
If we assume finally that we can make the substitution $\eta \rightarrow -iE$ 
in this expression, and that it corresponds to
$\eta^{\frac{k-1}{2}} \rightarrow |\eta|^{\frac{k-1}{2}}  
\rightarrow |E|^{\frac{k-1}{2}}$, we get:
\begin{equation}
\label{asym_reel} 
\rho(E) = \rho(0) - \alpha |E|^{\frac{k-1}{2}}+ o(E^{\frac{k-1}{2}}) \ ,
\end{equation}
a form in perfect agreement with the numerical determination of the density of
states for $k=2,3$ and $4$ (shown in Fig.~\ref{fig_cusp} for $k=2$). 
The justification of the jump from (\ref{asym_im}) to (\ref{asym_reel}) is not
obvious; indeed, the small $\eta$ 
expansion shows that $\rho$ is not analytic near $0$, hence no 'analytic 
continuation' argument can be straightforwardly applied. However as the 
density of states has to be invariant under $E+i\eta \rightarrow -E+i\eta$, 
one could assume for $\textrm{Im} z >0$ an Ansatz 
$\rho(z) = \rho(0)+f_1(|z|)+f_2(z)$ with $f_{1,2}$ regular and 
$f_2 = o(f_1)$ when $|z| \rightarrow 0$.

\section{From the moments of a random variable to its cumulative distribution function}
\label{app_moments}

In Sec.~\ref{sec_density_to_moments} we have seen how the asymptotic behavior
of a probability density close to the edge of its support translates into
the behavior of its large order moments. This appendix is devoted to
the reverse direction of this connection, namely to what can be learnt on a
probability measure from the asymptotic knowledge of its moments. In 
particular we provide the proofs of Eqs.~(\ref{eq_statement_integrated_density_od},\ref{eq_statement_integrated_density_d}).

Consider a non-negative random variable $X$, with a probability 
distribution $\eta$ supported on $[0,E_+]$, where $E_+$ is a priori unknown. 
Note that the assumption that $\eta$ is supported on non-negative reals is 
not restrictive for the cases we want to consider here: in 
the off-diagonal disorder case one can use the symmetry of $\rho$ to come
back to the above situation, and in the diagonal case the density of states
is supported on $[E_-,E_+]$ with $|E_-| < |E_+|$, hence the negatively 
supported part of the probability measure is asymptotically irrelevant. Thus Eqs. (\ref{eq_statement_integrated_density_od},\ref{eq_statement_integrated_density_d}) will be direct consequences of (\ref{eq_statement_offd},\ref{eq_statement_d}) and (\ref{eq_integrated_density}).

Let us decompose the expression of the $n$-th moment of $X$, for an arbitrary 
$\delta \in[0,E_+]$, as:
\beq 
u_n = \E[X^n] = \E[X^n | X< E_+-\delta] \P[X<E_+-\delta]+ 
\E[X^n | X\geq E_+-\delta] \P[X\geq E_+-\delta] \ .
\eeq
From that equation one easily obtains:
\beq 
\left(1-N(E_+-\delta) \right) (E_+-\delta)^n  \leq u_n \leq 
(E_+-\delta)^n + E_+^n (1-N(E_+-\delta)) \ .
\eeq
Hence one has, for arbitrary $n$:
\begin{align} \label{eq_appb_upb_1}
1-N(E_+-\delta) &\leq  (E_+-\delta)^{-n} u_n  \ , \\ 
 \label{eq_appb_lwb_1} 
1-N(E_+-\delta) &\geq  E_+^{-n} u_n - \left(1-\frac{\delta}{E_+}\right)^n \ .
\end{align}
As this is valid for any $\delta >0$, this proves our first point about the 
link between the support of the density of states and the exponential growth 
of its moments:
 \beq 
\ln E_+ = \lim_{n \rightarrow \infty} \frac{1}{n} \ln u_n \ ;
\eeq
a similar argument can be found in~\cite{mckay}.

In order to obtain some bounds on the behaviour of $N(E_+-\delta)$ 
one has to choose the value of $n$ in 
Eqs.~(\ref{eq_appb_upb_1},\ref{eq_appb_lwb_1}) in an optimal way, for a 
given value of $\delta$. As our control on $u_n$ is limited to large values of
$n$, the bounds on the cumulative distribution function shall be relevant
only for small values of $\delta$. The result will of course depend on the 
precise form of $u_n$, or more precisely on its large $n$ asymptotic. In the 
following we will assume that $u_n$ has the large $n$ behavior of the form
(\ref{eq_statement_offd},\ref{eq_statement_d}) of central interest here;
the reasoning is however more general and can be applied to reconstruct
the behavior of $N$ from $u_n$ in the three examples of 
Sec.~\ref{sec_density_to_moments}. We shall thus make the following assumption 
on the large $n$ behaviour of $u_n$ in order to treat within the same frame 
the off-diagonal and diagonal disorder cases:

\textit{Assume that for some $c,\epsilon, x >0$ and $n_0 \in \mathbb{N}$ it holds for
$n \geq n_0$ that
\begin{equation}
\label{eq_hypothesis_moments}
-(1+\epsilon) E_+ \frac{c}{(\ln n)^2} \leq 
\frac{1}{n} \ln u_n - \ln(E_+) \leq 
-  E_+ \frac{c}{(\ln n)^{2+x}} \ .
\end{equation}
Then for $\delta > 0$ small enough one has:}
\begin{equation}
\label{eq_integrated_density} 
e^{-e^{\sqrt{\frac{E_+^2 c (1+3 \epsilon)}{\delta}}}} \leq 1-N(E_+-\delta) \leq e^{-e^{\left({\frac{E_+^2 c }{\delta}}\right)^{\frac{1}{2+2x}}}} \ .
\end{equation}
We start with the right inequality. Using the upperbound on
$u_n$ from (\ref{eq_hypothesis_moments}), (\ref{eq_appb_upb_1}) reads 
for $n \geq n_0$ :
\beq \label{eq_appb_upb_2}  
1-N(E_+-\delta) \leq  \left( 1- \frac{\delta}{E_+} \right)^{-n}  
e^{-E_+ c \frac{n}{(\ln n)^{2+x}}} \ .
\eeq
Now we take: 
\beq 
n = \lfloor e^{\left( \frac{E_+^2 c}{\delta} \right)^{\frac{1}{2+x}}}\rfloor  
+ 1 \ .
\eeq
For $\delta$ small enough $n$ becomes larger than $n_0$ and one gets, 
replacing into (\ref{eq_appb_upb_2}):
\beq  
1-N(E_+-\delta)  \leq  
e^{-e^{\left( \frac{E_+^2 c}{\delta}\right)^{\frac{1}{2+2x}}}}  \ .
\eeq
The left inequality can be obtained in a similar way: this time 
(\ref{eq_appb_lwb_1}) may be rewritten, under assumption 
(\ref{eq_hypothesis_moments}) and for $n$ large enough:
\beq \label{eq_appb_lwb2} 
1-N(E_+-\delta) \geq  e^{-(1+\epsilon)E_+ c\frac{n}{(\ln n)^2}} 
- \left(1-\frac{\delta}{E_+}\right)^n \ .
\eeq
To obtain a valid bound we want the first term to be much larger than 
the second, that is:
\beq \label{eq_appb_condition} 
e^{-(1+\epsilon) E_+ c \frac{1}{(\ln n)^2}} > 
\left( 1- \frac{\delta}{E_+} \right) \ .
\eeq
Thus we take 
\beq 
n = \lfloor e^{\sqrt{\frac{E_+^2 c (1+ 2\epsilon)}{\delta}}} \rfloor 
\eeq
which satisfies (\ref{eq_appb_condition}) and gives, replacing into 
(\ref{eq_appb_lwb2}):
\beq  
1-N(E_+-\delta)  \geq  e^{-e^{\sqrt{\frac{E_+^2 c (1+3 \epsilon)}{\delta}}}}  
\ .
\eeq

\section{Proof of the recursion formula for the moments}
\label{app_resolvent}

In this appendix we give a formal proof of 
Eqs.~(\ref{eq_recursion_vraie},\ref{eq_recursion_cavite}), making use
of generating functions; we follow the standard notations in this field,
namely if $F(z,x)$ is a formal series,
$[z^n x^q]F(z,x)$ denotes the coefficient of its $z^n x^q$ term.

Let us first introduce, for a given realization of the disorder, the 
generating function $F(z)=\la 0| \frac{1}{\I-z H}|0\ra$, which is equal up to
a change of variables to the resolvent at the root of $\T_k$. We thus have
$u_n = \E[ [z^n] F(z) ]$. We decompose the operator $H$ according to
\begin{equation} 
H = V | 0 \rangle \langle 0| + \sum_{i=1}^{k+1} J_i \left( |i \rangle \langle 0 | + |0 \rangle \langle i | \right) + \sum_{i=1}^{k+1} \widetilde{H}_i \ ,
\end{equation}
where $\widetilde{H}_i$ acts only on the subtree rooted at $i$, one of the $k+1$
neighbors of the root. We introduce the generating functions for each of these
subtrees, $F_i(z) = \langle 0 | \frac{1}{\I-z \widetilde{H}_i} | 0 \rangle$, 
and use the resolvent identity between operators 
$\frac{1}{A}-\frac{1}{B} = \frac{1}{A}(B-A)\frac{1}{B}$ with 
$A = \I - z H$, 
$B = \I-z\left( V | 0 \rangle \langle 0 | + 
\sum_{i=1}^{k+1} \widetilde{H}_i \right)$, to get:
\begin{equation} 
\begin{split} 
F(z) &= \frac{1}{1-zV} + \left \langle 0 \left | \frac{z}{\I-z H}  
\sum_{i=1}^{k+1} J_i 
\left( | i \rangle \langle 0 | +  |0 \rangle \langle i | \right) 
\frac{1}{\I-z V | 0 \rangle \langle 0 | - z \sum_{i=1}^{k+1} \widetilde{H}_i} 
\right | 0 \right \rangle
\\ & = \frac{1}{1-zV} \left(1 + z \sum_{i=1}^{k+1} J_i \left \langle 0 \left | \frac{1}{\I-z H} \right | i \right \rangle \right) \ ,
\end{split} 
\end{equation}
where we used the fact that $\widetilde{H}_i | 0 \rangle = 0$. 
One can show in a similar
way that $\left \langle 0 \left | \frac{1}{\I-z H} \right | i \right \rangle 
= z J_i F(z) F_i(z)$, so that
\begin{equation} 
F(z) = \frac{1}{1-zV}  \left(1 + F(z) z^2 \sum_{i=1}^{k+1} J_i^2 F_i(z) \right) 
\Longrightarrow F(z) = \frac{1}{1-zV-z^2 \sum_{i=1}^{k+1} J_i^2 F_i(z)} \ .
\end{equation}
Let us now introduce a bivariate generating function, 
$F(z,x) = \frac{1}{1-x F(z)}$. As a consequence of the recursion equation on
$F(z)$, one obtains:
\begin{equation} 
\begin{split} 
F(z,x) &= \frac{ 1-zV-z^2 \sum_{i=1}^{k+1} J_i^2 F_i(z)}
{ 1-x-zV-z^2 \sum_{i=1}^{k+1} J_i^2 F_i(z)} 
\\ & = \frac{1}{1-x} \left(1-zV - z^2 \sum_{i=1}^{k+1} J_i^2 F_i(z) \right) 
\sum_{l=0}^\infty \left(\frac{zV + z^2 \sum_{i=1}^{k+1} J_i^2 F_i(z) }{1-x} 
\right)^l
\\ &= \frac{1}{1-x}+ \sum_{l=1}^\infty\frac{1}{(1-x)^l} 
\left(\frac{1}{1-x}-1 \right) \left({zV + z^2 \sum_{i=1}^{k+1} J_i^2 F_i(z) } 
\right)^l
\\ &= \frac{1}{1-x} + \sum_{\substack{p,s\ge 0 \\ p+s \geq 1}} 
\frac{x}{(1-x)^{s+p+1}} z^{2p+s} 
\sum_{\substack{p_1, \dots, p_{k+1} \ge 0 \\ p_1 + \dots + p_{k+1} =p}} 
{s+p \choose s, p_1, \dots, p_{k+1} } V^s 
\prod_{i=1}^{k+1} J_i^{2 p_i} \prod_{i=1}^{k+1} F_i(z)^{p_i} \ .
 \end{split}
\label{eq_Fzx}
\end{equation}
As $u_n = \E [ [z^n x] F(z,x)] $, we obtain for $n\ge 1$
\begin{equation}
u_n=\sum_{\substack{
s,p_1,m_1,\dots,p_{k+1},m_{k+1} \ge 0\\
s+m_1+2 p_1 + \dots+ m_{k+1}+2 p_{k+1}=n}} {s+p \choose s, p_1, \dots, p_{k+1}}  
\E[J^{2p_1}] \dots \E[J^{2 p_{k+1}}] \E[V^s] 
\prod_{i=1}^{k+1} \E[[z^{m_i}] F_i(z)^{p_i}] \ ,
\end{equation}
which proves Eq.~(\ref{eq_recursion_vraie}) with 
$u(n,q)=\E[[z^n] F_i(z)^q]=\E[[z^n x^q] F_i(z,x) ]$. In this last expression
we introduced the bivariate generating function $F_i(z,x)=\frac{1}{1-xF_i(z)}$.
The proof of Eq.~(\ref{eq_recursion_cavite}) follows exactly the same lines,
using $\tT_k$ instead of $\T_k$ (hence the root has only $k$ neighbors), and
extracting the coefficient of order $x^q$ in the equation corresponding to
(\ref{eq_Fzx}) thanks to the identity 
$[x^{q-1}] (1-x)^{-(s+p+1)} = {s+p+q-1 \choose q-1}$.

\section{Computation of $u(n,q)$ in the pure case}
\label{app_pure_case}

In this appendix we explain how to find the solution~(\ref{eq_unq_pure_case})
to the recursive equation~(\ref{eq_recursion_cavite}) in the pure case. 
Following the interpretation of $u(n,q)$ in terms of walks on 
$\tT_k \cup{(-1)}$ given in Sec.~\ref{sec_recursion_relation}, one realizes 
that in the pure case $u^0(n,q)$ is $k^{n/2}$ times the number of Dyck paths 
of length $n+2q$ that start and end at height $1$ and visit $q$ times 
the height $0$. Following the reflection principle discussed in~\cite{feller1},
this amounts to count the number of paths from $(0,0)$ to $(n,q)$ that reach 
the height $q$ for the first time at abscissa $n$, hence 
$u^0(n,q) = {n+q \choose n/2+q} \frac{q}{n+q} k^{n/2}$. 
In particular this gives back the well-known value of $\tilde{c}^0_n=u^0(n,1)$ 
as the Catalan number times $k^{n/2}$.

One can also check analytically that this expression for $u^0(n,q)$ is indeed
the solution of the recursion relation (\ref{eq_recursion_cavite}).
The simplest way to do so is to take benefit of the equivalence between
(\ref{eq_recursion_cavite}) and the equations on bivariate generating 
functions from which we derived (\ref{eq_recursion_cavite}) in 
App.~\ref{app_resolvent}. Simplifying them in the pure case, one realizes
that the verification boils down to prove:
\begin{equation}
{n+q \choose n/2+q} \frac{q}{n+q} k^{n/2} = 
[z^n] \left( \frac{1-\sqrt{1-4 k z^2}}{2 k z^2}  \right)^q \ .
\end{equation}
This equality can be proven~\cite{feller1} by checking it directly for 
$q=1$ and all $n$, and by showing that both sides $l_{n,q}$ and $r_{n,q}$ 
of the equation obeys the same recurrence equation,
\begin{equation} 
l_{n,q}=\frac{l_{n+2,q-1}-l_{n+2,q-2}}{k} \ , \qquad 
r_{n,q}=\frac{r_{n+2,q-1}-r_{n+2,q-2}}{k} \ .
\end{equation}

\section{Dyck and Motzkin paths of restricted height}
\label{app_lower_bound}

\begin{figure}[h]
\center
\includegraphics[height=1.8cm]{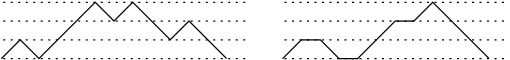}
\caption{Left: a Dyck path of length 12 and height 3.
Right: a Motzkin path of length 11 and height 3.}
\label{figure_dyck}
\end{figure}

We explain in this appendix the derivation of the properties of Motzkin
paths of restricted height that we used during the proofs of the lower bounds
in Sec.~\ref{sec_bounds_offd_lower} and~\ref{sec_bounds_d_lower}.
A Motzkin path of length $n$ is a sequence $h_0,h_1,\dots,h_n$ of 
non-negative integers such that $h_0=h_n=0$, 
$h_{i+1}-h_i \in \{+1,0,-1\}$ for all 
$i\in[0,n-1]$, see Fig.~\ref{figure_dyck} for an illustration. It is called
a Dyck path if there is no horizontal step $h_{i+1}=h_i$. Its height is the
maximum value of $h_i$ reached for $i\in[0,n]$. Let us assign a weight $a$ 
(resp. $c$, $1$) to an ascending (resp. horizontal, descending) step, and
define the weight of a Motzkin path as the product of the weight of its steps. 
We denote by $m_{n,h}$ the sum of these weights over the paths of length $n$ 
whose height is smaller or equal to $h$. The generating function 
$G^{(h)}(x)=\sum_{n \ge 0} m_{n,h} x^n$ is known to 
be~\cite{krattenthaler, flajolet}:
\begin{equation}
\label{eq_generating_function_depth_r} 
G^{(h)}(x) = \frac{1}{x} \frac{p_h(1/x)}{p_{h+1}(1/x)} \ ,
\end{equation}
where the $p_h(x)$ are polynomials of degree $h$, solution of the recurrence
equation:
\begin{equation} 
 p_{h+1}(x) =(x-c )\,p_h(x) - a \, p_{h-1}(x) \ , 
\hspace{1cm} \text{with} \ \ \ \ p_0(x) = 1 \ , \ \  p_1(x) = x- c \ . 
\end{equation}
It is convenient to change variables and define the polynomials $q_h$ by
$q_h(x) = \frac{1}{a^{r/2}} p_h(2 \sqrt{a}\, x+c)$. Indeed the recursion
becomes:
\begin{equation}
 q_{h+1}(x) =2x \, q_h(x) -  q_{h-1}(x) \ , 
\hspace{1cm} \text{with} \ \ \ \ q_0(x) = 1 \ , \ \  q_1(x) = 2x \ . 
\end{equation}
Hence one recognizes that $q_h(x) = \mathcal{U}_h(x)$, where 
$\mathcal{U}_h$ is the $h$-th Chebychev polynomial of the second kind. 

Consider now the rational function $\mathcal{U}_h(x)/\mathcal{U}_{h+1}(x)$; 
it has $h+1$ simple poles located at the roots of $\mathcal{U}_{h+1}(x)$, 
namely $x_j(h+1)=\cos\left(\frac{j\pi}{h+2} \right)$ with $j\in[1,h+1]$.
Using basic properties of the Chebychev polynomials one easily obtains
the following decomposition:
\begin{equation}
\frac{\mathcal{U}_h(x)}{\mathcal{U}_{h+1}(x)} = 
\sum_{j=1}^{h+1}\frac{1}{h+2} \sin^2\left(\frac{j\pi}{h+2} \right)
\frac{1}{x-x_j(h+1)} \ .
\end{equation}
Replacing the $p_h$ in terms of the Chebychev polynomials in 
Eq.~(\ref{eq_generating_function_depth_r}) and using the decomposition above
leads to an explicit expression of the generating function,
\begin{equation}
G^{(h)}(x) = \sum_{j=1}^{h+1}\frac{2}{h+2} \sin^2\left(\frac{j\pi}{h+2} \right)
\frac{1}{1-x \left(c+2 \sqrt{a}\cos\left(\frac{j\pi}{h+2} \right) \right)} \ ,
\end{equation}
and of the coefficient $m_{n,h}$ after the expansion in powers of $x$,
\begin{equation}
\label{eq_appendix_mnr}
m_{n,h} = \sum_{j=1}^{h+1}\frac{2}{h+2} \sin^2\left(\frac{j\pi}{h+2} \right)
 \left(c+2 \sqrt{a}\cos\left(\frac{j\pi}{h+2} \right) \right)^n \ .
\end{equation}
This expression is exact for all $n$ and $h$. For $n$ even all terms in the
sum are positive and one can lowerbound it by
retaining only the term $j=1$ and by using the inequality 
$\sin \theta \ge 2 \theta/\pi$ for $\theta \in [0,\pi/2]$. This yields
the bound
\begin{equation}
m_{n,h} \ge \left(\frac{2}{h+2}\right)^3 
\left(c+2 \sqrt{a}\cos\left(\frac{\pi}{h+2} \right) \right)^n \ ,
\end{equation}
which we used in Sec.~\ref{sec_bounds_offd_lower} 
with $a=1$ and $c=0$.

Let us now justify the asymptotic statement on $m_{n,h}$ used in 
Sec.~\ref{sec_bounds_d_lower} . Consider first the
behavior of $m_{n,h}$ for $c>0$, $n\to \infty$ with $h$ fixed. In
Eq.~(\ref{eq_appendix_mnr}) the term raised to the power $n$ is maximal for
$j=1$, and is strictly greater in absolute value than all others, hence
\begin{equation}
\lim_{n \to \infty} \frac{1}{n} \ln m_{n,h} = 
\ln\left(c + 2 \sqrt{a}\cos\left(\frac{\pi}{h+2} \right) \right) \ .
\label{eq_mnh_fixed}
\end{equation}
Suppose now that $h$ diverges in an $n$-dependent way. As long as 
$h\ll n^{1/2}$ one can check that the terms $j\ge 2$ are subdominant, and that
\begin{equation}
\frac{1}{n} \ln m_{n,h} = \ln(c+ 2\sqrt{a}) - 
\frac{\pi^2 \sqrt{a}}{c+2\sqrt{a}} \frac{1}{h^2} + O\left(\frac{1}{h^4}\right)
\ .
\label{eq_mnh_variable}
\end{equation}
This is the result used in Sec.~\ref{sec_bounds_d_lower} with $a=k$, $c=V_0$
and $h = \lfloor\alpha \ln n \rfloor$, which indeed satisfied the condition
$h\ll n^{1/2}$.

\section{The degeneracy of walks induced by self-bond steps}
\label{app_nstoone}
This appendix is devoted to an alternative proof a statement
used in Sec.~\ref{sec_bounds_d_upper}, namely that there are ${n \choose s}$
walks $\w$ of length $n$ with $s$ self-bond steps that share a common underlying
walk $\ow$ of length $n-s$, obtained by removing the self-bond steps from 
$\w$. Let us denote $\hsigma$ the common skeleton of length $n-s$.
From Eq.~(\ref{eq_moment_deplie})
one reads the combinatorial factor associated to such a skeleton complemented
by the numbers $\{s_v\}_{v \in \cV}$ of self-bond steps on each of the
vertices:
\begin{eqnarray}
\eta(\hsigma,\{s_v\}) &=& 
{s_0+ n_1 + n_2 + \dots +  n_{k+1} \choose s_0, n_1, n_2, \dots,  n_{k+1}}
\prod_{v \in \cV \setminus 0} 
{ n_{v} - 1 + s_v + n_{v_1} + \dots +  n_{v_k}  \choose n_v-1, s_v, n_{v_1}, \dots, n_{v_k}} \\
&=& \eta(\hsigma) {s_0+ n_1 + n_2 + \dots +  n_{k+1} \choose s_0}
\prod_{v \in \cV \setminus 0} 
{ n_{v} - 1 + s_v + n_{v_1} + \dots +  n_{v_k}  \choose s_v}
\end{eqnarray}
In the second line we have factored out the combinatorial factor
$\eta(\hsigma)$ of the walk deprived of the self-bond steps. We compute
now the sum over all possible choices of the $\{s_v\}_{v \in \cV}$:
\begin{eqnarray}
\sum_{\substack{\{s_v\}_{v \in \cV} \\ \sum_v s_v = s}} \eta(\hsigma,\{s_v\}) &=& 
\eta(\hsigma) [z^s] \left(
\sum_{s_0=0}^\infty {s_0+ n_1 + n_2 + \dots +  n_{k+1} \choose s_0} z^{s_0}\right)
\prod_{v \in \cV \setminus 0} \left( \sum_{s_v=0}^\infty
{ n_{v} - 1 + s_v + n_{v_1} + \dots +  n_{v_k}  \choose s_v} z^{s_v} \right) 
\nonumber \\
&=& \eta(\hsigma) [z^s] (1-z)^{-(1+n_1 + n_2 + \dots +  n_{k+1})}
\prod_{v \in \cV \setminus 0} (1-z)^{-(n_{v} + n_{v_1} + \dots +  n_{v_k})}
\label{eq_nstoone}\\ 
&=& \eta(\hsigma) [z^s] (1-z)^{-(1+ 2 \sum_{v \in \cV\setminus 0} n_v)} = 
\eta(\hsigma) {n \choose s} \ ,
\nonumber
\end{eqnarray}
where we have used the combinatorial identity 
$[z^p] (1-z)^{-(a+1)} = {p +a \choose p}$, recognized that for each edge $e$
$n_e$ appears twice (in the binomial coefficient associated to both
its endvertices), and noted that by definition of the length of the skeleton
$2 \sum_{v \in \cV \setminus 0} n_v = n-s$. Each of the $\eta(\hsigma)$ walks
$\ow$ without self-bond steps compatible with $\hsigma$ gives thus birth to
${n \choose s}$ walks with $s$ self-bond steps.

We shall now explain the proof of the upperbound (\ref{eq_ub_E}) on 
$E(\ow,s,\cA)$, the number of walks $\w \in \W_n$ that are obtained from 
$\ow \in \M_{n-s}$ by adding to it $s$ self-bond steps, in such a way that the 
self-support of $\w$ is $\cA$. This number is smaller than the number of 
walks $\w$ with a self-support included in $\cA$ (not necessarily equal to
$\cA$). This last quantity can be computed as
\begin{equation}
\frac{1}{\eta(\hsigma)} 
\sum_{\substack{\{s_v\}_{v \in \cA} \\ \sum_v s_v = s}} \eta(\hsigma,\{s_v\})
= {s-1 + \I(0 \in \cA)(1+n_1+\dots+n_{k+1}) + 
\underset{v \in \cA \setminus 0}{\sum}(n_v + n_{v_1}+\dots +n_{v_k}) 
\choose s} \ ,
\end{equation}
where we denoted $\I(M)$ the characteristic function of the event $M$, and
obtained the right-hand side with a reasoning similar to the one
in (\ref{eq_nstoone}). One can trade the sum over vertices for a sum over
the edges of the support by introducing the numbers $d_e(\cA) \in \{0,1,2\}$ 
which count the number of endvertices of the edge $e$ which belongs to $\cA$,
and rewrite the last quantity as
\begin{equation}
{n- (\I(0 \notin \cA) + \underset{e \in \sigma}{\sum} n_e(\ow) (2-d_e(\cA))) 
\choose s } \ .
\end{equation}
Finally (\ref{eq_ub_E}) is obtained by using the facts that $n_e(\ow) \ge 1$ 
for all edges in the support and that each vertex of $\cA$ has at most degree 
$k+1$ in the support.

\bibliographystyle{h-physrev}
\bibliography{biblio} 

\end{document}